\documentclass[runningheads]{llncs}
\usepackage[T1]{fontenc}
\usepackage{graphicx}
\usepackage{longtable,booktabs,array}
\usepackage{calc} % for calculating minipage widths
\usepackage{etoolbox}
\makeatletter
\patchcmd\longtable{\par}{\if@noskipsec\mbox{}\fi\par}{}{}
\makeatother
\IfFileExists{footnotehyper.sty}{\usepackage{footnotehyper}}{\usepackage{footnote}}
\makesavenoteenv{longtable}
\usepackage{graphicx}
\makeatletter
\def\maxwidth{\ifdim\Gin@nat@width>\linewidth\linewidth\else\Gin@nat@width\fi}
\def\maxheight{\ifdim\Gin@nat@height>\textheight\textheight\else\Gin@nat@height\fi}
\makeatother
\setkeys{Gin}{width=\maxwidth,height=\maxheight,keepaspectratio}
\makeatletter
\def\fps@figure{htbp}
\makeatother
\usepackage{lmodern}
\usepackage{mathtools} 
\usepackage{graphicx}
\graphicspath{{images/}}
\usepackage{subfig}
\usepackage{cite}
\usepackage{amsmath,amssymb,amsfonts}
\usepackage{algorithmic}
\usepackage{graphicx}
\usepackage{textcomp}
\usepackage{xcolor}
\usepackage{booktabs}
\usepackage{cleveref}
\usepackage{url} 
\def\BibTeX{{\rm B\kern-.05em{\sc i\kern-.025em b}\kern-.08em
    T\kern-.1667em\lower.7ex\hbox{E}\kern-.125emX}}

\usepackage[labelfont=bf]{caption}
\usepackage{subcaption}

\usepackage{ntheorem}
\theoremstyle{TH}
\newenvironment{appendixproof}[1]{\setlength{\parindent}{0pt}\paragraph{PROOF OF #1}}{\hfill$\blacksquare$\\}

    \makeatletter
    \newcommand{\mathleft}{\@fleqntrue\@mathmargin0pt}
    \newcommand{\mathcenter}{\@fleqnfalse}
    \makeatother

\begin{document}
\title{Optimal Threshold Signatures in Bitcoin}
% \author{}
% \institute{}
\author{Korok Ray\inst{1}\orcidID{0000-0001-8477-8079} \and
Sindura Saraswathi\inst{2}\orcidID{0009-0008-4491-5946}}

\authorrunning{K. Ray and S. Saraswathi}

\institute{Mays Business School, Texas A\&M University, College Station, TX, USA\\
\email{korok@tamu.edu}
\and
University of Central Florida, Orlando, FL, USA\\
\email{sindura@ucf.edu}\\
}
\maketitle              % typeset the header of the contribution
\begin{abstract}
We formulate the design of a threshold signature scheme as made possible on cryptocurrency protocols like Bitcoin. The funds are secured by an \textit{m-of-n} threshold signature, where at least $m$ signatures are needed to unlock the funds. A user designs this scheme knowing that a malicious attacker can also obtain the signatures with some probability. Higher thresholds offer more security, but also risk locking the user out of his own funds. The optimal threshold balances these twin effects. Interventions like increasing the security or usability of the signatures allow for higher thresholds. We model dynamic threshold signature schemes, where the probability of a user or attacker obtaining signatures changes with time. A dynamic threshold signature scheme is optimal, and increasing security or usability allows for higher thresholds and longer time locks. Together, these results show that threshold signature wallet design should be treated as an expected-loss optimization problem rather than a fixed convention such as 2-of-3 or 3-of-5.

\keywords{Threshold signature \and Security \and Usability \and Dynamic threshold signature}
\end{abstract}

\section{Introduction}\label{introduction}

Threshold signatures are now common methods to secure funds in
cryptocurrencies like Bitcoin. An \textit{m-of-n} signature scheme
requires at least $m$ signatures to unlock funds secured by
$n$ public keys. Threshold signatures offer the benefit of reducing
reliance on a single signature and the possibility of distributing
signatures to different people or locations. Bitcoin companies like
Unchained, Casa, and OnRamp sell products and services based on a
\textit{2-of-3} threshold signature setup.

An open question, however, is the optimal design of such schemes. When
should the threshold be high, when should it be low, and how should it
change over time? We study this problem in a security context, where a
malicious attacker tries to steal a user's signatures.
The user selects a signature threshold that trades off two forces.
A higher threshold will have the benefit of preventing an attacker from
accessing the funds, but it has the cost of also locking the user out of
his own funds. Self-lockout is a common problem, evidenced by the vast
sums that users are willing to pay to recover lost bitcoin. An optimal threshold
signature will balance these costs and benefits, namely,
the benefit of security against the cost of usability.

The user seeks to minimize his expected loss, which consists of two
components. The first is an attacker loss, which is the loss from an
attacker stealing his funds, and the second is the
user's loss of access to his own funds. We examine the
optimal threshold signature in a static model with respect to
security or usability. An example of increasing security would be
placing signatures in a safe, while increasing usability could involve
embossing private keys onto a titanium plate that would give the user
access to his signatures in case of a fire. We find that increasing
security or usability allows for higher optimal thresholds. This might
seem counterintuitive since, for example, placing signatures in a safe
might induce the user to relax their threshold. But greater security
decreases the probability of an attack, thereby making a high threshold
relatively cheaper, so the user can afford to adopt a higher threshold
than before.

Next, we formulate a dynamic model where the probabilities of the user
and attacker accessing the signatures decay over time. Over very long
horizons, users can lose access to private keys through loss or
negligence. We formulate a dynamic model where the user selects a
threshold in each stage and also a length of time (a time lock) when the
threshold changes. We solve for this optimal dynamic threshold signature
scheme and find that it does, in fact, optimally degrade, where the
early-stage threshold exceeds the later stage threshold. We also find
that the optimal time lock increases in security and in usability.

Conversely, we also consider a dynamic model in which the probability of the user accessing the signatures decays over time, while the probability of an attacker gaining access increases as the signatures are used more frequently. We solve for the optimal dynamic threshold signature scheme under these conditions and find that the threshold typically degrades over time, similar to the previous case. However, when the attacker’s gain rate significantly exceeds the user’s decay rate, the optimal strategy shifts, and the threshold is adjusted upward in later stages to preserve security while balancing usability.

The main contribution of this paper is to shift the design question from how to implement threshold signatures to how to choose them. Existing work largely studies cryptographic construction, security definitions, or attacks on threshold signing protocols; we instead study the wallet-design problem faced by a Bitcoin user choosing a threshold and timelock policy under theft and self-lockout loss. This gives a practical design rule that optimal custody requires minimizing expected loss across attacker access and user access.

\subsection{Application to Bitcoin}\label{application-to-bitcoin}

Bitcoin is the oldest and largest cryptocurrency as measured by market
capitalization. Launched in 2008 by the pseudonymous creator, Satoshi
Nakamoto, Bitcoin is a secure and immutable way to store and transfer
value across a global network of computer nodes backed by the
proof-of-work algorithm. The Bitcoin whitepaper \cite{Nakamoto2008} lays out the
basic design, and several textbooks, such as Song \cite{Song2019}, Narayanan
\cite{NarayananEtAl2016}, and Antonopoulos \cite{Antonopoulos2017} provide in-depth expositions of the protocol. 

The original implementation of Bitcoin allowed for a simple
implementation of threshold signature, in which more than one signature is
needed to form a bitcoin transaction. An intuitive example of a
threshold signature in the real world would be a nuclear submarine that
might require authorization from both the first officer and the captain
before a launch. More generally, a threshold signature transaction has
$m$ of $n$ spending conditions, where there are $n$
public keys and at least $m$ signatures needed to unlock the
Bitcoin. Threshold signatures can be useful to prevent the risk of theft and
loss from a single signature. For example, under a $2/3$ threshold signature
scheme, only two signatures are required to transfer the bitcoin. So, if
the user loses one signature, he can still recover his bitcoin.

The original design of threshold signature was functional but somewhat
clunky, as it revealed the spending conditions: the number of signatures
and public keys, since all of these were elements in the stack needed
for script evaluation. The last major update to Bitcoin, Taproot,
provided much greater privacy, as it hid all of the spending conditions
inside a binary tree only whose root would appear on the blockchain. In
so doing, viewers of the blockchain could not differentiate between a
Taproot tree, called a Taptree, that was a simple address (called a
pay-to-Taproot address), or a complex script.

Dynamic threshold signature has a special implementation in Bitcoin that the
2021 Taproot upgrade enabled. Taproot allows bitcoin transactions to
contain complex scripts represented within large Merkle trees. These
trees can contain multiple signatures and time locks, making an on-chain
footprint no larger than a standard spending transaction.

\subsection{Related Work}\label{related-work}

Several studies have explored threshold signature schemes, with numerous
new proposals emerging in the literature. Many studies have also focused
on evaluating potential attacks or threats, as well as addressing the
security aspects of threshold signatures.

\cite{GoldfederEtAl2014} proposes the implementation of threshold signature scheme
compatible with Bitcoin's ECDSA signatures to improve
Bitcoin security. \cite{VanDerMerwe2007} introduces a new threshold signature
approach that eliminates the need for a trusted third party, leveraging
publicly verifiable DKG protocol with optimal rounds. \cite{DasEtAl2023} highlights
that existing threshold signatures with compact signatures and constant
verification times are not suitable when signers have different weights.
It introduces a novel threshold signature scheme that relies on an
inner-product argument, supporting arbitrary weight distributions and
thresholds using 8 group elements and requiring 8 exponentiations and
bilinear pairings for verification. \cite{DasRen2024}, \cite{KomloGoldberg2021}, \cite{DoernerEtAl2024} and
\cite{EspitauEtAl2024} also present new approaches to threshold signature scheme. In
contrast, our work focuses on determining the optimal threshold for an
existing threshold signature scheme --- an area that has not been
sufficiently explored. This optimization is crucial to enhancing the
efficiency of threshold signature schemes, setting our work apart by
addressing a critical gap in existing research.

According to \cite{TakeiShudo2024}, distributing signing power among M participants
out of N shares provides redundancy and prevents a small group from
having full control over the signing process. However, if N equals M, it
increases availability risks, and if N is much larger than M, it poses a
threat to confidentiality. \cite{AumassonShlomovits2020} discusses real-world deployment and
attack models on threshold signature scheme implementations used
by major organizations, identifying three attacks that exploit software
vulnerabilities and could potentially compromise assets like cold
wallets. One-more unforgeability (OMUF) in \cite{NavotTessaro2025}, aims to improve
security for threshold signatures by ensuring that an adversary can only
create a limited number of signatures using unique, non-reused signature
shares. \cite{FischlinEtAl2025} introduces Beyond UnForgeability Features (BUFF) for
threshold signature schemes, which protect against attacks involving
maliciously chosen keys, such as stealing a message-signature pair. It
formalizes these features for threshold schemes and presents a generic
compiler that adds exclusive ownership and message-bound properties with
minimal overhead. \cite{BellareEtAl2022} presents a security framework for
non-interactive threshold signature schemes, including a hierarchy of
increasingly stronger security definitions. While these works address
threats, attacks, and security, the novelty of our model lies in its
focus on adapting the threshold signature scheme to improve both
security and usability for the user. We also propose a novel model to
determine the optimal threshold and optimal timelocks for a dynamic
signature scheme, which introduces a new dimension of efficiency
and flexibility.

\section{The Model}\label{the-model}

A risk-neutral decision maker seeks to design an optimal threshold signature
scheme to custody his bitcoin. Suppose such a scheme is an \textit{m-of-n}
threshold signature where $n$ is the number of public keys and $m$ is the
minimal number of signatures needed to sign the transaction. In what
follows, refer to signing the transaction, unlocking the bitcoin, and
spending the funds as equivalent ways to describe the successful process
of executing the multi-signature script. Throughout, we refer to the
user and the attacker as having access to the signatures, even though
the common convention is access to the underlying private keys that
generate those signatures.

Let \(\tau\) denote the multi-signature threshold as a fraction between
0 and 1. So for an \textit{m of n} threshold signature scheme,
\(\tau = m / n\). Common choices of threshold signature schemes used in
practice are 2-of-3 or 3-of-5, which generates a threshold of
\(\tau = 2 / 3\) or \(\tau = 3 / 5\), respectively. We reduce the
problem of selecting the optimal \textit{m of n} threshold signature scheme by
selecting the optimal threshold $\tau$. By assumption, we treat different
threshold signature schemes that generate the same threshold as equivalent.
So, a 2-of-3 scheme is formally equivalent to a 4-of-6 scheme, since
they have the same threshold. In practice, different threshold signature
schemes with the same threshold but different $m$ and $n$ numbers may be
qualitatively different, but we leave that to future research.

A malicious attacker has some possible access to the
user's signatures. If the attacker has more signatures
than the threshold, he will then be able to unlock the
user's bitcoin and steal the funds. The core trade-off
for the user is to select a threshold that guarantees both security and
usability. High thresholds are more secure because they place a larger
hurdle on the attacker to obtain more signatures, but they are also less
usable because they require the user to manage more of these multiple
signatures. To formalize this, let \(p(\tau)\) be the probability that
the user has more than \(\tau\) signatures and \(q(\tau)\) be the
probability that the attacker has more than \(\tau\) signatures. We are
agnostic on which specific signatures the attacker obtains, since what
matters for the threshold requirement is the number of signatures rather
than which signatures. So, for a 3-of-5 threshold signature scheme, it does
not matter if the attacker has signatures 1, 2, and 4 or 3, 4, and 5. Either case will be sufficient to clear the threshold.

Because of the threshold nature of the threshold signature scheme, both the
attacker and the user can unlock the bitcoin if and only if he has more
than the threshold number of signatures. Let
\(\theta = (\theta_{u},\theta_{a})\) be the state space, where
\(\theta_{u} = 1\) if the user has more than $\tau$ signatures, and
\(\theta_{a} = 1\) if the attacker has more than \(\tau\) signatures.
There are, therefore, four states of the world given by
\(\{(0,0),(0,1),(1,0),(1,1)\}\). Assume that the probabilities \(p\) and
\(q\) are independent.\footnote{In practice, there may be some correlation between \(p\) and \(q\). For example, if the attacker knows something about the user's behavior or about some population of users that are more or less likely to lose their private keys and therefore their signatures. But here, we take the more standard assumption that the attacker does not have this information.}

In each state of the world, we can also define the loss function to the
user. Suppose the transaction has value V, which is the current market
value of the bitcoin that the threshold signature scheme is securing. If the
user has more than $\tau$ signatures but the attacker does not, then there is
no loss. If instead, the attacker has access to more than \(\tau\)
signatures but the user does not, there is a full loss of V. If neither
has access to the threshold number of signatures, then neither the
attacker nor the user can access the funds, so there is still a loss of
V. Finally, if the attacker and user both have more than the threshold
number of signatures, then we assume there is an even chance that there
will be a loss of funds since we can resolve these ties with a coin
flip.

\begin{figure}[t]
    \centering
    \includegraphics[width=1\linewidth]{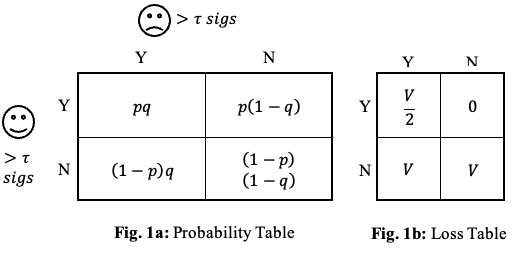}
    \caption{Probability and Loss Tables}
    \label{fig:2}
\end{figure}

\Cref{fig:2} shows the probability table and the loss function for each of
the four states of the world. The probability table emerges from the
independence of the two probabilities. The loss table shows the private
loss to the user in each of these states of the world, depending on
whether he or the attacker has more than the threshold number of
signatures. As is standard with security analysis, we focus on the
private loss, rather than the social loss. A computation of social loss
would put some nonzero weight on the attacker's welfare.
If we were to attach a nonzero weight to the attacker receiving the
funds, then the social loss would be smaller than the private loss
because the funds would transfer from the user to the attacker. However,
our base assumption is that we analyze this from the user's perspective,
and therefore his payoffs are relevant, rather than the joint payoffs.

% \begin{figure}[t]
%     \centering
%     \includegraphics[width=1\linewidth]{image5.jpg}
%     \caption{Equivalence classes of threshold signatures}
%     \label{fig:1}
% \end{figure}

\subsection{The Optimal Threshold}\label{the-optimal-threshold}

The key decision problem for the user is selecting the threshold
\(\tau\), which determines the ratio \(\frac{m}{n}\) in an \textit{m-of-n}
threshold signature. We assume that the probabilities \(p\) and \(q\) are
both continuously decreasing functions of the threshold \(\tau\). This
captures the intuition that lower thresholds are more usable because the
user can access the funds with fewer signatures, but they are also less
secure because the attacker can obtain more of the signatures. This is
the core representation of the trade-off between usability and security.
Said differently, higher thresholds are more secure because they require
the attacker to obtain more signatures, but they are also less
convenient for the user because they require the user to manage and
maintain these signatures.

As a simple example, a 1-of-5 scheme requires the user to only have a
single signature to spend his funds, but it also allows an attacker to
obtain only a single signature, leading to a high attack surface. A
5-of-5 signature requires an attacker to have all five signatures and
therefore is highly secure, but it is also inconvenient for the user
because losing any of the signatures would lock him out from his own
funds. Future research can develop a more elaborate behavioral model for
the full shape of these probability functions. For now, we take these as
more reduced-form representations of a broader decision problem.

The user will therefore select the threshold to minimize the expected
loss across all four states of the world. Therefore, the user will solve

\begin{equation}
    \min_{\tau\  \in \ \lbrack 0,\ 1\rbrack}{\frac{V}{2}p(\tau)q(\tau) + V\left( 1 - p(\tau) \right)}
\end{equation}

The expected loss in the objective function above is the sum of two
terms. The first term is the expected loss from an attack, which we call
the ``attack loss'' for shorthand. This occurs if the user and attacker
both have more than the threshold number of signatures, so there is an
even chance that the user will lose his funds. The second term is the
expected loss from the user no longer having a sufficient number of
signatures, in which case he is locked out of his own funds. We call
this the ``user loss'' for shorthand. Note that the user loss is
independent of whether the attacker has the necessary number of
signatures.

To generate closed-form solutions, assume that the probabilities \(p\)
and \(q\) are given by

\begin{equation}
    p(\tau) = 1 - \frac{a}{k}\tau^{k}\ \text{ and }\ q(\tau) = 1 - \frac{b}{k}\tau^{k}
\end{equation}

Observe that both probabilities are 1 at a threshold of \(\tau = 0\) and
that both probabilities continuously decrease in the level of the
threshold. If \(k = 2\), the first derivative of both probabilities is
given by \(p^{\prime}(\tau) = \  - a\tau\) and \(q^{\prime}(\tau) = - b\tau\),
while the second derivative is given by \(p^{\prime \prime}(\tau) = \  - a\) and
\(q^{\prime \prime}(\tau) = - b\). 
% The parameters \(a\) and \(b\) track the
% curvature of the probability functions. To see this, consider Figure 3,
% which shows the probability functions for different levels of \(a\) and
% \(b\). 
The parameters \(a\) and \(b\) track the curvature of the
probability functions \(p\) and \(q\). They measure how responsive the
probability is to a marginal change in the threshold of the
multi-signature scheme. A high parameter for \(a\) or \(b\) will mean
that an increase in the threshold will lead to a larger decrease in the
probability of the user or attacker having access to the funds.

To see an example, consider a shift from a 3-of-5 scheme to a 4-of-5
scheme. The higher threshold means the probability of the user or
attacker accessing the funds decreases. If the user loses a signature,
there is now a higher chance that he will get locked out of his funds,
whereas if the attacker obtains a signature, there is a lower chance
that the attacker will be able to access the funds (because the
threshold is higher). So, the probabilities will fall for both \(p\) and
\(q\) when shifting from a 3-of-5 to a 4-of-5, but different design
choices from the user can affect how much these probabilities change.

For example, suppose the user shifts some or all of the signatures onto
a titanium, fire-resistant material. This would be an example of a
decrease in \(a\), but no change in \(b\); it would not affect the
attacker's ability to obtain the signatures, but would
improve the user's access to the signatures in the case
of a fire. Similarly, placing these signatures into a locked safe rather
than an unlocked drawer could decrease \(b\) with no change in \(a\),
since accessing a safe with a four-digit pin is not onerous for the
user, but could hinder an attacker. Finally, there may be some choices
that increase both \(a\) and \(b\), such as moving some signatures
off-site, like a safe deposit box, or buried in a yard. This would
decrease the chance of an attacker finding it, but also increase the
chance that the user cannot access this signature when he needs to. Our
next result solves for the optimal threshold under these functional
forms.

\begin{proposition}\label{prop1}
    With probability functions \(p(\tau) = 1 - \frac{a}{2}\tau^{2}\) and
\(q(\tau) = 1 - \frac{b}{2}\tau^{2}\), the optimal threshold signature
scheme is

\begin{equation}
    \tau^{*} = \sqrt{\frac{b - a}{ab}}
\end{equation}
\end{proposition}

\Cref{prop1} shows that the optimal threshold is a function of the
difference between \(b\) and \(a\), the parameters governing the
curvature of the probability functions \(q\) and \(p\), respectively. To
gain intuition on this result, observe that the user seeks an
environment where the probability that he can access his funds is high,
while the probability that an attacker can access the funds is low. This
will occur with a high \(p\) and a low \(q\). In terms of the parameters, this happens with a low \(a\) and a high \(b\). So, the user prefers a large difference between \(b\) and \(a\). If the user
can arrange this difference, then he can afford to adopt a high
threshold, because the high threshold will forestall the attack but not
decrease usability for himself. If the user cannot arrange a large gap
between \(b\) and \(a\), then the reverse is true. Then, the user must
settle for a smaller threshold, which will increase the probability of
an attack, but simultaneously decrease the chance that the user is
locked out of his own funds by losing one of his signatures.

% \addtocounter{figure}{+1}
% \begin{figure}[t]
%     \centering
%     \includegraphics[width=1\linewidth]{image1.png}
%     \caption{Probability Functions}
%     \label{fig:3}
% \end{figure}

Another way to see this is that threshold signatures impose both a
benefit and a cost to the user and the attacker. A high threshold has
the benefit of locking out the attacker, but the cost of also locking
out the user. When \(b\) is high, the benefit exceeds the cost, and the
user can afford a higher threshold. When \(b\) is low, the cost exceeds
the benefit, and the user must settle for a lower threshold. The
sufficient condition is that \(b > a\). So, the attacker's probability
function must be more sensitive to changes in the threshold than the
user's probability function to guarantee that the
threshold is positive.

We can now conduct some straightforward comparative statics on the
optimal threshold to discover what happens from a change in the
user's environment via changes in \(b\) and \(a\).

\begin{corollary}\label{cor1}
    Increasing security or usability leads to higher
optimal thresholds.
\end{corollary}

To use a prior example, increasing security by moving from a paper
wallet to a hardware wallet will lead to a increase in \(b\) and
therefore a higher threshold. Similarly, fortifying and converting and
embedding seed phrases onto a titanium plate that is fire-resistant
would be an increase in usability or a decrease in \(a\) which would
also lead to a higher optimal threshold.

\begin{figure}[t]
    \centering
    \includegraphics[width=1\linewidth]{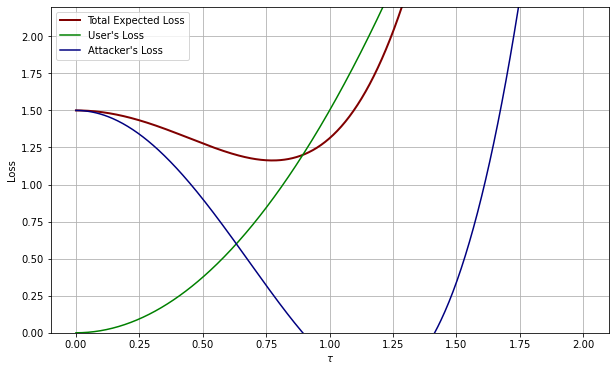}
    \caption{The attacker’s, user’s, and total expected loss functions}
    \label{fig:4}
\end{figure}

\Cref{fig:4} plots the expected loss, user loss, and attacker loss for
\(a = 1\) and \(b = 2.5\). First, observe that the user loss is
invariant with respect to $b$, and increases in the threshold. This
reflects the simple logic that higher thresholds are more likely to lock
a user out of his funds. The attacker loss function is more complex. It
is decreasing for thresholds in our domain of interest, namely,
thresholds between 0 and 1. Therefore, increasing the threshold leads to
a lower probability of an attack, since there is less of a likelihood
that the attacker will have the necessary number of signatures.

The expected loss function is the sum of the user loss and the attacker
loss. Increasing the threshold has a first-order effect of decreasing
the attacker loss function, because the attacker is less likely to have
the necessary number of signatures. Increasing the threshold will also
lock the user out of his funds. This has a moderating effect on the loss
function. And so, the attacker loss function is initially decreasing and
concave - it decreases but at a decreasing rate\footnote{\textsuperscript{}
  This can be seen more explicitly by noticing that the attacker loss
  function, in fact, eventually reaches a
  global minimum at a threshold greater than one. While this is not
  economically relevant, it does explain why the expected loss function,
  which is the sum of the two loss functions, also has the similar shape.} -- but eventually becomes convex as the
threshold increases.

To understand \Cref{cor1}, refer again to \Cref{fig:4}.
% which plots the
% total expected loss function, the user loss, the attacker loss, and the
% total loss. 
Decreasing the probability of an attack by increasing b has
no effect on the user's loss. But it does shift the
attacker loss function downward. Because the expected loss is the sum of
the user's loss and the attacker's loss, the decrease in the
attacker's loss pulls down the expected loss function.
This function has a higher minimal point, leading to a greater optimal
threshold that minimizes the expected loss.

Decreasing the user's probability of obtaining the
necessary threshold signatures by increasing \(a\) has two effects.
First, it increases the user's expected loss by pulling
the curve upward, since \(a\) is the slope of the user's
loss function. Second, it decreases the attacker's loss
function. The net effect is that the expected loss, the combination of
two loss functions, shifts up and to the left. So the optimal threshold
that minimizes the total expected loss is now smaller.

The interaction between the combination of the two loss functions
explains why the model predicts results that may at first conflict with
standard intuition. If the user places his private keys into a safe that
deters a potential thief, one might expect that the user can relax his
threshold signature since he has increased the security of his
setup. But as the prior discussion shows, decreasing the
attacker's probability (by increasing \(b\)) lowers the
loss from an attack, and this consequently lowers the total expected
loss, allowing the user to afford a higher threshold. In effect, the
lower probability of theft makes a high threshold ``cheaper'', which is
why the user should increase, rather than decrease, the threshold.
Similarly, placing private keys onto a fire-resistant plate will
increase the probability of the user retaining his signatures,
equivalent to a decrease in \(a\). This will lower both the
user's loss from an attack, and the
attacker's loss. Both effects lead to a lower total
expected loss, which is the combination of the two. This lower expected
loss makes high thresholds cheaper, so the user should optimally pick a
higher, not lower, threshold number of signatures.

\section{Dynamic threshold signature}\label{dynamic-multisig}

Over time, the probability of maintaining keys changes. In particular,
users can lose their private keys over time, while attackers may either see their ability to launch successful attacks erode or, conversely, improve their capability if users compromise or mishandle keys as time progresses. As such, we can examine
how threshold signature schemes should change over time. This is
particularly relevant for Bitcoin because of the time lock feature.
Taproot, the last major upgrade to Bitcoin in 2021, allows much richer
smart contracts on Bitcoin using large Merkle trees that can be up to
128 levels deep, holding up to \(2^{128}\) objects. These leaves in the
tree can be either data or scripts. In particular, they can contain time
locks and dynamic threshold signature schemes, where the spending conditions
change over time.

\begin{figure}[t]
    \centering
    \includegraphics[width=0.7\linewidth]{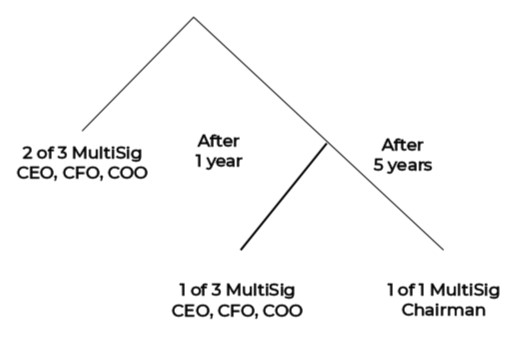}
    \caption{An example of a dynamic threshold signature.}
    \label{fig:5}
\end{figure}

To see a simple example, consider \Cref{fig:5}, where a company that holds
Bitcoin in its treasury and has a 2-of-3 threshold signature scheme split
between the CEO, CFO, and COO. A workaround is to build a
threshold signature transaction that automatically degrades after some time
interval, such as one year. At that point, the scheme could degrade into
a 1-of-3 threshold signature scheme to prevent loss of funds if two of the
three executives lose their keys. After five years, the scheme could
degrade even further into a 1-of-1 scheme to a different user, such as
the chairman. This could cover contingencies if all three keys are lost,
such as in a plane crash or fire.

We seek to model an optimal dynamic threshold signature scheme and analyze
its theoretical properties. The core assumption is that the probability
of obtaining greater than the threshold number of signatures will decay
over time, reflecting the behavior described above. We then seek to know
if a dynamic threshold signature scheme performs better than one without
degradation, and the properties of such an optimal scheme.

To model this formally, suppose that the probability functions \(p\) and
\(q\) are subject to exponential time decay. Now, let the probability
functions be

\begin{equation}
    p(\tau)e^{- \lambda t} \ \text{ and } \ q(\tau)e^{- \gamma t}
\end{equation}

where \(\lambda\) is the rate of decay for the user, and \(\gamma\) is
the rate of decay for the attacker. The definitions of these
probabilities are the same as before, namely, they reflect the
probability that either the attacker or user has more than the threshold
number of signatures needed to unlock the funds. As before, these are
statistically independent probabilities. The state space is identical,
with \(\theta_{u} = 1\) if the user has more than the threshold number
of signatures, and similarly for \(\theta_{a} = 1\). So, at any point in
time, the probability table and loss table are the same as before.

As a benchmark, we consider the first program with a benchmark (i.e.
time-invariant) threshold signature scheme under these decaying
probabilities. In this benchmark, we will minimize the expected loss,
where the expectation now is taken not only over the state space, but
also over time. The benchmark program solves the following problem:

\begin{align}
    \min_{\tau}\frac{V}{2}\int_{0}^{\infty}&{p(\tau)e^{- \lambda t}q(\tau)e^{- \gamma t}dt} \\
    &+ V\int_{0}^{\infty}{\left( 1 - p(\tau)e^{- \lambda t} \right)dt} \nonumber
\end{align}

We propose an \textit{n}-stage dynamic threshold signature scheme given by
\(\left( \tau_{i},T_{i} \right)\). This is a time interval \(T_{i}\),
which conveys the point where the threshold signature scheme changes. For
all time \(t \in \left( T_{i - 1},T_{i} \right)\), the scheme uses a
threshold \(\tau_{i}\), where \(T_{o} = 0\) and \(T_{n} = \infty\). A
dynamic threshold signature, therefore, solves the following program:

\begin{equation}\label{eq:dynamic_optimization}
    \begin{aligned}
    \min_{\left( \tau_{i},\tau_{i + 1},T \right)} \sum_{i = 1}^{n} \left \{
    \frac{V}{2}\int_{T_{i}}{p\left( \tau_{i} \right)e^{- \lambda t} q\left( \tau_{i} \right) e^{- \gamma t}dt} \right.\\
    \left. + V\int_{T_{i}}{\left( 1 - p\left( \tau_{i} \right)e^{- \lambda t} \right)dt} \right\}
    \end{aligned}
\end{equation}
    
A dynamic threshold signature performs better than a time-invariant threshold signature,
because the dynamic program allows the user to tailor thresholds per
stage.

We now seek to solve for the optimal dynamic threshold signature contract,
for \(p(\tau) = 1 - \frac{a}{2}\tau^{2}\) and
\(q(\tau) = 1 - \frac{b}{2}\tau^{2}\).

We can then solve this program to characterize the optimal dynamic
threshold signature contract.

\begin{proposition}\label{prop2}
    The optimal dynamic threshold signature contract is given by

    \begin{align}\label{eq:timelock}
        \tau_{i}^{*} &= \sqrt{\frac{b + a\left( 1 - \frac{{2y}_{i}}{Z_{i}} \right)}{ab}} \\
        T^{*} &= \frac{1}{\gamma} \ln\left\lbrack \frac{p \left( \tau_{i - 1}^{*} \right)q\left( \tau_{i - 1}^{*} \right) - p\left( \tau_{i}^{*} \right)q\left( \tau_{i}^{*} \right)}{2\left( p\left( \tau_{i - 1}^{*} \right) - p\left( \tau_{i}^{*} \right) \right)} \right\rbrack
    \end{align}

    \noindent where

    \begin{equation}
        Z_{i} = \int_{T_{i}}^{}{e^{- \lambda t}e^{- \gamma t}dt}
    \end{equation}

    \noindent and

    \begin{equation}
        y_{i} = \int_{T_{i}}^{}{e^{- \lambda t}dt}
    \end{equation}
    
    \noindent for \(i = 1,..,n\).
\end{proposition}

The optimal scheme has several noteworthy properties. First, observe
that if there is no time decay for the attacker, then \(\gamma = 0\),
and therefore \(y_{i} = Z_{i}\), and so the optimal dynamic threshold signature is
identical to the optimal benchmark threshold signature. Therefore, the user losing
access to his keys over time is not sufficient to justify a dynamic
threshold signature. A compelling justification requires that the attacker's
capabilities also vary over time - either decreasing
\((q(\tau)e^{- \gamma t})\) or increasing (\(q(\tau)e^{\gamma t})\).

The latter scenario is particularly relevant: as a user continues to
utilize their keys over time, the risk of key compromise grows due to
increased exposure, thereby enhancing the attacker's ability to mount a
successful attack. We formalize this case in the following proposition.

\begin{proposition}\label{prop3}
    The optimal dynamic threshold signature contract is
given by

\begin{equation}
    \tau_{i}^{*} = \sqrt{\frac{b + a\left( 1 - \frac{{2y}_{i}}{Z_{i}} \right)}{ab}}
\end{equation}

\begin{equation}\label{eq:timelock2}
    T^{*} = \frac{1}{\gamma}\ln\left\lbrack \frac{2\left( p\left( \tau_{i - 1}^{*} \right) - p\left( \tau_{i}^{*} \right) \right)}{p\left( \tau_{i - 1}^{*} \right)q\left( \tau_{i - 1}^{*} \right) - p\left( \tau_{i}^{*} \right)q\left( \tau_{i}^{*} \right)} \right\rbrack
\end{equation}

where

\begin{equation}
    Z_{i} = \int_{T_{i}}^{}{e^{- \lambda t}e^{\gamma t}dt}
\end{equation}

and

\begin{equation}
    y_{i} = \int_{T_{i}}^{}{e^{- \lambda t}dt}
\end{equation}

for \(i = 1,..,n\).
\end{proposition}

The next corollary shows that the optimal threshold signature scheme over
time does in fact degrade when both the user's and the attacker's abilities of gaining access to signatures decrease.
\begin{corollary}\label{cor2}
    The optimal dynamic threshold signature scheme degrades
\(\left( \tau_{i - 1}^{*} > \tau_{i}^{*} \right)\).
\end{corollary}

Recall that the optimal threshold balances two twin effects. The benefit
of a higher threshold is that it forestalls malicious attacks, but the
cost is that it can also lock you out of your own funds. As the
probability of obtaining a signature for either the attacker or the user
decays over time, so too does the cost of a high threshold. As time
marches on, the probability of an attacker obtaining a signature
decreases, so the user can lower the threshold. It is
interesting that \Cref{cor2} does not depend on the specific rates of
decay of the two probabilities, \(\lambda\) or \(\gamma\), 
nor does it
depend on the curvature of the probability functions of the attacker or
user, \(a\) or \(b\). Regardless of those parameter levels, it is always
optimal to degrade the threshold signature configuration. This gives a theoretical
foundation for the example discussed earlier, where a 4-of-5
threshold signature decays to a 2-of-5 after some time. Now,
let's see how the optimal time lock changes with respect
to the parameters of interest:

With this result, we can now see how the optimal timelock changes with
respect to the probabilities of obtaining signatures.

\begin{corollary}\label{cor3}
    Increasing security or usability allows for higher
thresholds and longer timelocks. So $\frac{\partial\tau_{i}}{\partial a} < 0,\frac{{\partial\tau}_{i}}{\partial b} > 0,\frac{{\partial T}^{*}}{\partial a} < 0$ and $\frac{{\partial T}^{*}}{\partial b} > 0$.
\end{corollary}

\Cref{cor3} shows that the optimal time lock decreases in \(a\) but
increases in \(b\). To gain intuition here, notice that, as we mentioned
earlier, decreasing the attacker's probability (by
increasing b) lowers the loss from an attack, and this consequently
lowers the total expected loss, allowing the user to afford a higher
threshold. In effect, the lower probability of theft makes a high
threshold ``cheaper'', which is why the user should increase, rather
than decrease, the time under the threshold. If instead the parameter
\(a\) decreases, this is equivalent to an increase in the probability
function \(p\), so the user is more likely to retain his signatures. The
user can, therefore, afford to operate in the high threshold stage, and
is willing to expand the time lock.

\begin{corollary}\label{cor4}
    The optimal threshold signature \(\tau_{i}^{*}\) decreases in \(\lambda\), and decreases in \(\gamma\).
\end{corollary}

\begin{corollary}\label{cor5}
    If $f$ satisfies the MLRP with respect to $g$, the optimal threshold signature timelock $T^{*}$ decreases in \(\lambda\). The optimal threshold signature timelock $T^{*}$ decreases in \(\gamma\) as well.
\end{corollary}
The effects of the time decay parameter has an equivalent effect on the
optimal time lock. For example, if \(\lambda\) increases, the time decay
increases, so the user is less likely to retain his signatures over
time. The user's expected loss is now higher, making a high threshold
expensive. In response, the user can decrease both the threshold and the
time under that threshold (the timelock).

The next corollary shows that the optimal threshold signature scheme over time when the attacker’s ability of gaining access
to signatures increases.

\begin{corollary}\label{cor6}
With attacker gaining access over time, the optimal dynamic threshold signature scheme degrades
\(\left( \tau_{i - 1}^{*} > \tau_{i}^{*} \right)\) when $0<\gamma<\lambda$. Conversely, \(\left( \tau_{i - 1}^{*} < \tau_{i}^{*} \right)\) when $\gamma > \lambda$.
\end{corollary}

\begin{corollary}\label{cor7}
    With attacker gaining access over time
(\(q(\tau)e^{\gamma t})\),
\(\frac{\partial\tau_{i}}{\partial a} > 0,\frac{{\partial\tau}_{i}}{\partial b} < 0\)\emph{,}\(\frac{{\partial T}^{*}}{\partial a} > 0\)
and \(\frac{{\partial T}^{*}}{\partial b} < 0\), unless
\(\gamma \gg \lambda\).
\end{corollary}

\begin{corollary}\label{cor8}
    With attacker gaining access over time (\(q(\tau)e^{\gamma t})\), the optimal threshold signature \(\tau_{i}^{*}\) increases in \(\lambda\), and increases in \(\gamma\).
\end{corollary}

\begin{corollary}\label{cor9}
    With attacker gaining access over time \(q(\tau)e^{\gamma t})\), if $f$ satisfies the MLRP with respect to $g$, the optimal threshold signature timelock $T^{*}$ increases in \(\lambda\). The optimal threshold signature timelock $T^{*}$ increases in \(\gamma\) as well if \(\frac{f^{\prime}(\gamma)}{f(\gamma)} - \frac{g^{\prime}(\gamma)}{g(\gamma)} > T^{*}\).
\end{corollary}

\section{Simulation}\label{simulation}

\begin{figure}[t]
    \centering
    \includegraphics[width=0.7\linewidth]{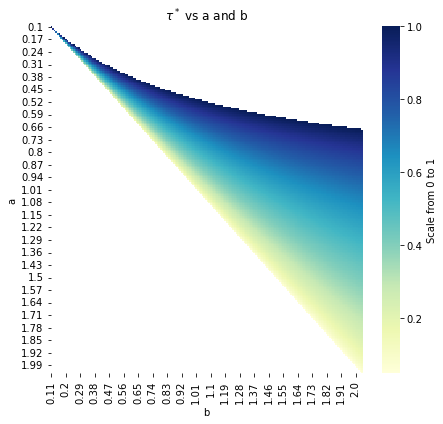}
    \caption{Relationship between $\tau^{*}$, $a$, and $b$ (\Cref{prop1}}
    \label{fig:fig1}
\end{figure}

\begin{figure}[t]
    \centering
    \includegraphics[width=1\linewidth]{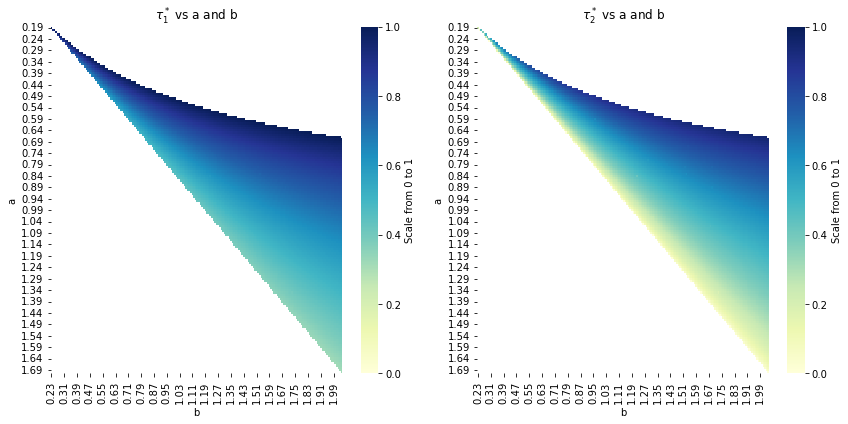}
    \caption{Relationship between $\tau^{*}_{1}$, $\tau^{*}_{2}$, $a$, and $b$ (\Cref{prop2})}
    \label{fig:fig2}
\end{figure}

\begin{figure}[t]
    \centering
    \includegraphics[width=1\linewidth]{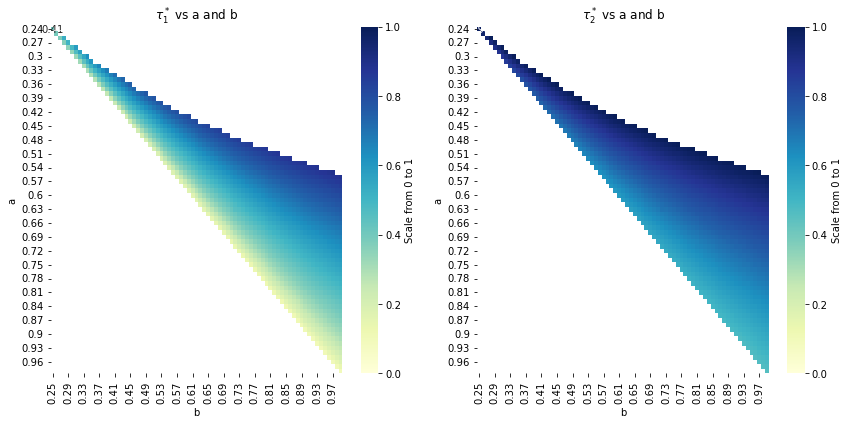}
    \caption{Relationship between $\tau^{*}_{1}$, $\tau^{*}_{2}$, $a$, and $b$ (\Cref{prop3})}
    \label{fig:fig3}
\end{figure}
We ran simulations to explore \Cref{prop1}, \Cref{prop2} and \Cref{prop3}. Our simulations run through a range of values for parameters $a$ and $b$. The values of $a$ and $b$ are iterated in such a way that $b$ is always greater than $a$, to avoid
redundant calculations and focus on meaningful combinations. The goal of
our simulation is to find the optimal benchmark threshold value
\(\tau^{*}\) and 2-stage dynamic threshold values  \( \tau_{1}^{*}\ and\ \tau_{2}^{*}\). The results of the simulation are visualized through heatmaps to observe
how the optimal thresholds
\(\tau^{*},\ \tau_{1}^{*}\ and\ \tau_{2}^{*}\) change with respect to
the values of $a$ and $b$ following propositions. 

From \Cref{fig:fig1}, we observe the relationship between $a$, $b$, and the optimal threshold $\tau^{*}$. Specifically, as $b$ increases or $a$ decreases, the value of $\tau^{*}$ increases. \Cref{fig:fig2} illustrates how $\tau_{1}^{*}$ and $\tau_{2}^{*}$ vary with $a$ and $b$ under different combinations of $\gamma$ and $\lambda$, assuming that both the user's and attacker's ability to access signatures decay over time. It is evident from the figure that, for any given values of $a$ and $b$, $\tau_{1}^{*}$ is consistently greater than $\tau_{2}^{*}$, providing further insights into how the optimal thresholds adapt with respect to the optimal time lock $T^{*}$ as specified in \eqref{eq:timelock}. These findings are consistent with the result stated in \Cref{cor2}.

Similarly, \Cref{fig:fig3} depicts how $\tau_{1}^{*}$ and $\tau_{2}^{*}$ adapt to changes in security and usability parameters when the user gradually loses access to signatures while the attacker's ability to access signatures improves over time, modeled as $q(\tau)e^{\gamma t}$. In this scenario, $\gamma$ is chosen to be greater than $\lambda$. As shown, $\tau_{2}^{*}$ exceeds $\tau_{1}^{*}$, consistent with \Cref{cor6}. Intuitively, this increase in the threshold is necessary to enhance security, as a higher attacker gain rate ($\gamma$), resulting from more frequent usage by the user, increases the likelihood of signature compromise.

We implemented code using Python3 to construct and visualize Taproot trees incorporating complex spending conditions, such as a dynamic threshold signature scheme. Additionally, we developed a function to simulate Pay-to-Taproot (P2TR) transactions and broadcast them on the Bitcoin testnet. These implementations leverage libraries provided in \cite{key}. 

% The source code for our simulations is available on Github at: \url{https://github.com/sindurasaraswathi/Optimal_Threshold_Signatures}

\section{Conclusion}\label{conclusion}
The threshold in a Bitcoin custody scheme is not just an implementation parameter, but an economic security choice. A higher threshold reduces theft loss but increases self-lockout loss, so the optimal policy must balance both forces. In dynamic settings, this balance naturally leads to threshold schedules and timelocks where the wallet should harden or degrade access requirements depending on how user-access and attacker-access probabilities evolve over time. Taproot makes these policies practically relevant because complex threshold and timelock conditions can be represented using Taptrees.

% Threshold signature have been a part of Bitcoin since its
% inception. While Bitcoin's scripting language, Bitcoin
% Script, is limited by design to limit Bitcoin's attack
% surface, it has always been able to handle the control flow for multiple
% signatures and multiple public keys. Taproot was a significant advance
% because it improved the privacy of bitcoin transactions while at the
% same time increasing their complexity. Our analysis here of optimal
% threshold signature helps to guide some of the wide range of possibilities
% that Taproot unlocks. Indeed, the Taproot Merkle trees (Taptrees) can
% have vast depth and complexity, and therefore, some theory is useful in
% terms of guiding which schemes are superior to others.

Future research can extend this line of work by broadening the economic
and security environment of future users of threshold signature schemes. The
undiscovered country will be unlocking the full possibility of Taproot
to allow for all manner of economic transactions to be represented
through Taptrees and more complex contracts, including time locks and
multiple signatures. Even greater possibilities will emerge when
second-layer protocols integrate with the base layer. This can be
particularly useful for artificial agents to utilize the full complexity
that Taproot enables. While this paper has focused chiefly on the human
use cases of threshold signature, future users may be AI agents
who can utilize Bitcoin Script in ways we cannot yet imagine.

% \bibliographystyle{custom-unsrt}
% \bibliography{reference}
\bibliographystyle{unsrt}
\bibliography{OMSAI_FC_26}

@misc{Nakamoto2008,
  author       = {Satoshi Nakamoto},
  title        = {Bitcoin: A Peer-to-Peer Electronic Cash System},
  year         = {2008},
  howpublished = {\url{https://bitcoin.org/bitcoin.pdf}},
  note         = {White paper}
}

@book{Song2019,
  author    = {Jimmy Song},
  title     = {Programming Bitcoin: Learn How to Program Bitcoin from Scratch},
  year      = {2019},
  publisher = {O'Reilly Media},
  address   = {Sebastopol, CA}
}

@book{NarayananEtAl2016,
  author    = {Arvind Narayanan and Joseph Bonneau and Edward Felten and Andrew Miller and Steven Goldfeder},
  title     = {Bitcoin and Cryptocurrency Technologies: A Comprehensive Introduction},
  year      = {2016},
  publisher = {Princeton University Press},
  address   = {Princeton, NJ}
}

@book{Antonopoulos2017,
  author    = {Andreas M. Antonopoulos},
  title     = {Mastering Bitcoin: Programming the Open Blockchain},
  edition   = {2},
  year      = {2017},
  publisher = {O'Reilly Media},
  address   = {Sebastopol, CA}
}

@inproceedings{GoldfederEtAl2014,
  author    = {Steven Goldfeder and Joseph Bonneau and Joshua A. Kroll and Edward W. Felten},
  title     = {Securing Bitcoin Wallets via Threshold Signatures},
  booktitle = {Workshop on Bitcoin and Blockchain Research (Financial Cryptography 2014)},
  year      = {2014}
}

@article{VanDerMerwe2007,
  author  = {J. Van Der Merwe and D. S. Dawoud and S. McDonald},
  title   = {A Fully Distributed Proactively Secure Threshold‑Multisignature Scheme},
  journal = {IEEE Transactions on Parallel and Distributed Systems},
  volume  = {18},
  number  = {4},
  pages   = {562--575},
  year    = {2007},
  month   = apr,
  doi     = {10.1109/TPDS.2007.1005}
}

@inproceedings{DasEtAl2023,
  author    = {Sourav Das and Philippe Camacho and Zhuolun Xiang and Javier Nieto and Benedikt B{\"u}nz and Ling Ren},
  title     = {Threshold Signatures from Inner Product Argument: Succinct, Weighted, and Multi‑threshold},
  booktitle = {Proceedings of the 2023 ACM SIGSAC Conference on Computer and Communications Security},
  pages     = {356--370},
  year      = {2023}
}

@incollection{DasRen2024,
  author    = {Sourav Das and Ling Ren},
  title     = {Adaptively Secure BLS Threshold Signatures from DDH and co‑CDH},
  booktitle = {International Cryptology Conference ( CRYPTO 2024 )},
  pages     = {251--284},
  publisher = {Springer Nature Switzerland},
  address   = {Cham},
  year      = {2024}
}

@incollection{KomloGoldberg2021,
  author    = {Chelsea Komlo and Ian Goldberg},
  title     = {FROST: Flexible Round‑Optimized Schnorr Threshold Signatures},
  booktitle = {Selected Areas in Cryptography (SAC 2020): 27th International Conference, Revised Selected Papers},
  pages     = {34--65},
  publisher = {Springer International Publishing},
  year      = {2021}
}

@inproceedings{DoernerEtAl2024,
  author    = {Jack Doerner and Yashvanth Kondi and Eysa Lee and Abhi Shelat},
  title     = {Threshold ECDSA in Three Rounds},
  booktitle = {Proceedings of the 2024 IEEE Symposium on Security and Privacy (SP)},
  pages     = {3053--3071},
  publisher = {IEEE},
  year      = {2024}
}

@incollection{EspitauEtAl2024,
  author    = {Thomas Espitau and Guilhem Niot and Thomas Prest},
  title     = {Flood and Submerse: Distributed Key Generation and Robust Threshold Signature from Lattices},
  booktitle = {International Cryptology Conference ( CRYPTO 2024 )},
  pages     = {425--458},
  publisher = {Springer Nature Switzerland},
  address   = {Cham},
  year      = {2024}
}

@inproceedings{TakeiShudo2024,
  author    = {Yuki Takei and Kazuyuki Shudo},
  title     = {Pragmatic Analysis of Key Management for Cryptocurrency Custodians},
  booktitle = {2024 IEEE International Conference on Blockchain and Cryptocurrency (ICBC)},
  address   = {Dublin, Ireland},
  pages     = {747--765},
  year      = {2024},
  doi       = {10.1109/ICBC59979.2024.10634356}
}

@misc{AumassonShlomovits2020,
  author       = {Jean‑Philippe Aumasson and Omer Shlomovits},
  title        = {Attacking Threshold Wallets},
  year         = {2020},
  howpublished = {Cryptology ePrint Archive: Report 2020/xxx}
}

@incollection{NavotTessaro2025,
  author    = {Sela Navot and Stefano Tessaro},
  title     = {One‑More Unforgeability for Multi‑ and Threshold Signatures},
  booktitle = {International Conference on the Theory and Application of Cryptology and Information Security (ASIACRYPT 2025)},
  pages     = {429--462},
  publisher = {Springer},
  address   = {Singapore},
  year      = {2025}
}

@misc{FischlinEtAl2025,
  author       = {Marc Fischlin and Aikaterini Mitrokotsa and Jenit Tomy},
  title        = {BUFFing Threshold Signature Schemes},
  year         = {2025},
  howpublished = {Cryptology ePrint Archive: Report 2025/xxx}
}

@incollection{BellareEtAl2022,
  author    = {Mihir Bellare and Elizabeth Crites and Chelsea Komlo and Mary Maller and Stefano Tessaro and Chenzhi Zhu},
  title     = {Better than Advertised Security for Non‑Interactive Threshold Signatures},
  booktitle = {International Cryptology Conference ( CRYPTO 2022 )},
  pages     = {517--550},
  publisher = {Springer Nature Switzerland},
  address   = {Cham},
  year      = {2022}
}

@misc{key,
	author = {Jimmy Song},
	title = {Programming-Taproot},
	howpublished = {\url{https://programmingbitcoin.com/programming-taproot/}},
	year = {},
	note = {},
}

\appendix
\section{Appendix}

\begin{appendixproof}{PROPOSITION~\ref{prop1}}
    The first and second-order conditions from the user's optimization

\begin{equation}\label{FOC}\tag{$FOC$}
    \frac{V}{2}\left\lbrack p^{\prime}(\tau)q(\tau) + {p(\tau)q}^{\prime}(\tau) \right\rbrack - {Vp}^{\prime}(\tau) = 0
\end{equation}

\vspace{20pt}
and

\begin{align}
    \frac{V}{2}&\left[ \left( p^{\prime \prime}(\tau)q(\tau) + p^{\prime}(\tau)q^{\prime}(\tau) \right) \right. \nonumber    \\
    &\left.+ \left( p^{\prime}{(\tau)q}^{\prime}(\tau) + {p(\tau)q}^{\prime \prime}(\tau) \right) \right\rbrack \nonumber \\
    &- {Vp}^{\prime \prime}(\tau) > 0 \label{SOSC}\tag{$SOSC$}
\end{align}

The first-order condition is a necessary condition that will generate
the optimal threshold for the user's minimized expected loss function.
The second-order condition is, as is, a sufficient condition for the
loss function to be convex to guarantee that the optimal threshold
occurs at a minimum rather than a maximum. Plugging the probability
functions \(p^{\prime}(\tau) = \  - a\tau\) and \(q^{\prime}(\tau) = - b\tau\)
into (\ref{FOC}) generates

\begin{equation}
    \frac{V}{2}\left\lbrack( - a\tau)\left(1 - \frac{b}{2}\tau^{2}\right) + \left(1 - \frac{a}{2}\tau^{2}\ \right)( - b\tau)\right\rbrack + Va\tau = 0
\end{equation}

Rearranging terms, this is

\begin{equation}
    V\left( ab\tau^{2} - a - b \right) + 2Va = 0
\end{equation}

Solving for the threshold \(\tau^{*}\) gives

\begin{equation}
    \tau^{*} = \sqrt{\frac{b - a}{ab}}
\end{equation}

Plugging the probability functions \(p^{\prime}(\tau) = \  - a\tau\),
\(q^{\prime}(\tau) = - b\tau\), \(p^{\prime \prime}(\tau) = \  - a\), and
\(q^{\prime \prime}(\tau) = - b\) into (\ref{SOSC}):

\begin{equation}
    \begin{aligned}
        \frac{V}{2}\left\lbrack - a\left( 1 - \frac{b}{2}\tau^{2} \right) + 2ab\tau^{2} + \left( 1 - \frac{a}{2}\tau^{2} \right)( - b) \right\rbrack +Va > 0
    \end{aligned}
\end{equation}

Rearranging terms, this is equivalent to

\begin{equation}
    \frac{V}{2}\left\lbrack 3ab\tau^{2} - b - a \right\rbrack + Va > 0
\end{equation}

We need the left hand side to be positive, simplifying to

\begin{equation}
    3ab\tau^{2} > b - a
\end{equation}

Plugging in the optimal threshold \(\tau^{*}\) shows that this equation
reduces to

\begin{equation}
    \frac{(b - a)(3ab)}{ab} + a - b > 0,
\end{equation}

which is always true.

\end{appendixproof}

\begin{appendixproof}{COROLLARY~\ref{cor1}}
    We interpret increases in security as increases in
\(b\). So, the probability function decreases by more for a given change
in the threshold. Differentiating the optimal threshold with respect to
\(b\), it is clear that \(\frac{{\partial\tau}^{*}}{\partial b} > 0\).
Similarly, increases in usability correspond to decreases in \(a\) and
\(\frac{{\partial\tau}^{*}}{\partial a} < 0\).
\end{appendixproof}

\begin{appendixproof}{PROPOSITION~\ref{prop2}}
    We can write \eqref{eq:dynamic_optimization} as:
    \begin{equation}
    \begin{aligned}
        \min_{\tau_{i}} \sum_{i = 1}^{n} \left\{ \frac{V}{2}Z_{i}p\left( \tau_{i} \right)q\left( \tau_{i} \right) \right.\\
        \left. + V\left( 1 - y_{i}p\left( \tau_{i} \right) \right) \right\}
    \end{aligned}
    \end{equation}

Differentiating with respect to the threshold in stage $i$ gives

\begin{equation}
    \frac{V}{2}Z_{i}\left( p^{\prime}\left( \tau_{i} \right)q\left( \tau_{i} \right) + q^{\prime}\left( \tau_{i} \right)p\left( \tau_{i} \right) \right) - {Vy}_{i}p^{\prime}\left( \tau_{i} \right) = 0
\end{equation}

We know that

\begin{equation}
    p\left( \tau_{i} \right) = 1 - \frac{a}{2}{\tau_{i}}^{2}\ \text{ and }\ q\left( \tau_{i} \right) = 1 - \frac{b}{2}{\tau_{i}}^{2}
\end{equation}

So we calculate that

\begin{align}
    p^{\prime}(\tau) &= \  - a\tau_{i} \\
    q^{\prime}(\tau) &= - b\tau_{i}
\end{align}

Substituting in these functions gives

\begin{align}
    \frac{V}{2}Z_{i}\left( - a\tau_{i}\left(1 - \frac{b}{2}{\tau_{i}}^{2}\right) + - b\tau_{i}\left(1 - \frac{a}{2}{\tau_{i}}^{2}\right) \right)& \nonumber \\
    - {Vy}_{i}( - a\tau_{i}) &= 0 \\
    \left( ab\tau_{i}^{2} - a - b \right)\frac{Z_{i}}{y_{i}} - 2a &= 0
\end{align}

Rearranging terms and solving gives

\begin{equation}
    \tau_{i}^{*} = \sqrt{\frac{b + a\left( 1 - \frac{{2y}_{i}}{Z_{i}} \right)}{ab}}
\end{equation}

Let \(T_{i - 1}\  = \ (T_{i - 1},T_{i})\) and
\(T_{i} = \ (T_{i},T_{i + 1})\). Let

\begin{align}
    X_{i} &= \frac{V}{2}\int_{T_{i}}^{}{p(\tau)q(\tau)e^{- \lambda t}e^{- \gamma t}dt} \\
    &\text{\hspace{50pt}and} \nonumber \\
    Y_{i} &= V\int_{T_{i}}^{}{\left( 1 - p(\tau)e^{- \lambda t} \right)dt}
\end{align}

(22) becomes

\begin{equation}
    \max_{\left( \tau_{i}\ ,\ \ \ T \right)}{X_{i - 1} + Y_{i - 1} +}X_{i} + Y_{i}
\end{equation}

Now, differentiate with respect to the time cutoff \(T\) to get:

\begin{align}
    \frac{{\partial X}_{i - 1}}{\partial T} &= \frac{V}{2}\left\lbrack p\left( \tau_{i - 1} \right)q\left( \tau_{i - 1} \right)e^{- \lambda T}e^{- \gamma T} \right\rbrack \\
    \frac{{\partial X}_{i}}{\partial T} &= \frac{V}{2}\left\lbrack - p\left( \tau_{i} \right)q\left( \tau_{i} \right)e^{- \lambda T}e^{- \gamma T} \right\rbrack \\
    \frac{{\partial Y}_{i - 1}}{\partial T} &= V\left\lbrack 1 - p\left( \tau_{i - 1} \right)e^{- \lambda T} \right\rbrack \\
    \frac{{\partial Y}_{i}}{\partial T} &= V\left\lbrack - \left( 1 - p\left( \tau_{i} \right)e^{- \lambda T} \right) \right\rbrack
\end{align}

Setting the sum of these to zero, we have

\begin{equation}
    \begin{aligned}
        \frac{V}{2}e^{- \lambda T}e^{- \gamma T}\left\lbrack p\left( \tau_{i - 1} \right)q\left( \tau_{i - 1} \right) - p\left( \tau_{i} \right)q\left( \tau_{i} \right) \right\rbrack& \\
        + {Ve}^{- \lambda T}\left\lbrack p\left( \tau_{i} \right) - p\left( \tau_{i - 1} \right) \right\rbrack &= 0
    \end{aligned}
\end{equation}

Rearranging terms, we have

\begin{align}
    e^{- \gamma T}&\left\lbrack p\left( \tau_{i - 1} \right)q\left( \tau_{i - 1} \right) - p\left( \tau_{i} \right)q\left( \tau_{i} \right) \right\rbrack \nonumber \\
    &= 2\left( p\left( \tau_{i - 1} \right) - p\left( \tau_{i} \right) \right) \\
    - \gamma T &= ln\left\lbrack \frac{2\left( p\left( \tau_{i - 1} \right) - p\left( \tau_{i} \right) \right)}{p\left( \tau_{i - 1} \right)q\left( \tau_{i - 1} \right) - p\left( \tau_{i} \right)q\left( \tau_{i} \right)} \right\rbrack \\
    T^{*} &= \frac{1}{\gamma}\ln\left\lbrack \frac{p\left( \tau_{i - 1}^{*} \right)q\left( \tau_{i - 1}^{*} \right) - p\left( \tau_{i}^{*} \right)q\left( \tau_{i}^{*} \right)}{2\left( p\left( \tau_{i - 1}^{*} \right) - p\left( \tau_{i}^{*} \right) \right)} \right\rbrack
\end{align}

at the optimal contract
\(\left( \tau_{i - 1}^{*},\ \tau_{i}^{*},T^{*} \right)\).

\end{appendixproof}

\begin{appendixproof}{PROPOSITION~\ref{prop3}}
Let

\begin{equation}
    Z_{i} = \int_{T_{i}}^{}{e^{- \lambda t}e^{\gamma t}dt}
\end{equation}

and

\begin{equation}
    y_{i} = \int_{T_{i}}^{}{e^{- \lambda t}dt}
\end{equation}

Let \(T_{i - 1}\  = \ (T_{i - 1},T_{i})\) and
\(T_{i} = \ (T_{i},T_{i + 1})\). Let

\begin{align}
    X_{i} &= \frac{V}{2}\int_{T_{i}}^{}{p(\tau)q(\tau)e^{- \lambda t}e^{\gamma t}dt} \\
    &\text{\hspace{50pt}and} \nonumber \\
    Y_{i} &= V\int_{T_{i}}^{}{\left( 1 - p(\tau)e^{- \lambda t} \right)dt}
\end{align}

(22) becomes

\begin{equation}
    \max_{\left( \tau_{i}\ ,\ \ \ T \right)}{X_{i - 1} + Y_{i - 1} +}X_{i} + Y_{i}
\end{equation}

Now, differentiate with respect to the time cutoff \(T\) to get:

\begin{align}
    \frac{{\partial X}_{i - 1}}{\partial T} &= \frac{V}{2}\left\lbrack p\left( \tau_{i - 1} \right)q\left( \tau_{i - 1} \right)e^{- \lambda T}e^{\gamma T} \right\rbrack \\
    \frac{{\partial X}_{i}}{\partial T} &= \frac{V}{2}\left\lbrack - p\left( \tau_{i} \right)q\left( \tau_{i} \right)e^{- \lambda T}e^{\gamma T} \right\rbrack \\
    \frac{{\partial Y}_{i - 1}}{\partial T} &= V\left\lbrack 1 - p\left( \tau_{i - 1} \right)e^{- \lambda T} \right\rbrack \\
    \frac{{\partial Y}_{i}}{\partial T} &= V\left\lbrack - \left( 1 - p\left( \tau_{i} \right)e^{- \lambda T} \right) \right\rbrack
\end{align}

Setting the sum of these to zero, we have

\begin{equation}
    \begin{aligned}
        \frac{V}{2}e^{- \lambda T}e^{\gamma T}\left\lbrack p\left( \tau_{i - 1} \right)q\left( \tau_{i - 1} \right) - p\left( \tau_{i} \right)q\left( \tau_{i} \right) \right\rbrack \\
        + {Ve}^{- \lambda T}\left\lbrack p\left( \tau_{i} \right) - p\left( \tau_{i - 1} \right) \right\rbrack = 0
    \end{aligned}
\end{equation}

Rearranging terms, we have

\begin{equation}
    \begin{aligned}
        e^{\gamma T}\left\lbrack p\left( \tau_{i - 1} \right)q\left( \tau_{i - 1} \right) - p\left( \tau_{i} \right)q\left( \tau_{i} \right) \right\rbrack \\
        = 2\left( p\left( \tau_{i - 1} \right) - p\left( \tau_{i} \right) \right)
    \end{aligned}
\end{equation}

\begin{align}
    \gamma T &= ln\left\lbrack \frac{2\left( p\left( \tau_{i - 1} \right) - p\left( \tau_{i} \right) \right)}{p\left( \tau_{i - 1} \right)q\left( \tau_{i - 1} \right) - p\left( \tau_{i} \right)q\left( \tau_{i} \right)} \right\rbrack \\
    T^{*} &= \frac{1}{\gamma}\ln\left\lbrack \frac{2\left( p\left( \tau_{i - 1} \right) - p\left( \tau_{i} \right) \right)}{p\left( \tau_{i - 1} \right)q\left( \tau_{i - 1} \right) - p\left( \tau_{i} \right)q\left( \tau_{i} \right)} \right\rbrack
\end{align}

at the optimal contract
\(\left( \tau_{i - 1}^{*},\ \tau_{i}^{*},T^{*} \right)\).

\end{appendixproof}

\begin{appendixproof}{COROLLARY~\ref{cor2}}
We wish to prove that, in the general case with n stages of degradation
with \(T_{0} = 0\) and \(T_{n} = \infty\),
\(\tau_{i - 1}^{*} > \tau_{i}^{*}\). We begin this proof by plugging in
\(\tau_{i}^{*} = \sqrt{\frac{b + a\left( 1 - \frac{{2y}_{i}}{Z_{i}} \right)}{ab}}\).

\begin{equation}
    \sqrt{\frac{b + a\left( 1 - \frac{{2y}_{i - 1}}{Z_{i - 1}} \right)}{ab}} > \sqrt{\frac{b + a\left( 1 - \frac{{2y}_{i}}{Z_{i}} \right)}{ab}}
\end{equation}

Cancelling all the similarities by squaring both sides, multiplying by
$ab$, subtracting $b$, dividing by $a$, and subtracting 1
from both sides leaves us with

\begin{equation}
    - \frac{{2y}_{i - 1}}{Z_{i - 1}} > - \frac{{2y}_{i}}{Z_{i}}
\end{equation}

Now, we divide by -2 and reverse the comparison to reach the equation

\begin{equation}
    \frac{y_{i - 1}}{Z_{i - 1}} < \frac{y_{i}}{Z_{i}}
\end{equation}

We need to find the conditions under which this equation is true.

First, we substitute in the equations for
\(y_{i - 1},\ y_{i},\ Z_{i - 1},Z_{i}\ \) to get

\begin{equation}
    \frac{\int_{T_{i - 2}}^{T_{i - 1}}{e^{- \lambda t}dt}}{\int_{T_{i - 2}}^{T_{i - 1}}{e^{- \lambda t}e^{- \gamma t}dt}} < \frac{\int_{T_{i - 1}}^{T_{i}}{e^{- \lambda t}dt}}{\int_{T_{i - 1}}^{T_{i}}{e^{- \lambda t}e^{- \gamma t}dt}}
\end{equation}

Integrating all parts gives us the equation

\begin{equation}
    \begin{aligned}
        &\frac{\frac{1}{\lambda}(e^{- \lambda T_{i - 2}} - e^{- \lambda T_{i - 1}})}{\frac{1}{\lambda + \gamma}(e^{- (\lambda + \gamma)T_{i - 2}} - e^{- (\lambda + \gamma)T_{i - 1}})} \\
        &\hspace{10pt} < \frac{\frac{1}{\lambda}(e^{- \lambda T_{i - 1}} - e^{- \lambda T_{i}})}{\frac{1}{\lambda + \gamma}(e^{- (\lambda + \gamma)T_{i - 1}} - e^{- (\lambda + \gamma)T_{i}})}
    \end{aligned}
\end{equation}

Canceling out \(\frac{1}{\lambda}\) and \(\frac{1}{\lambda + \gamma}\)
and cross multiplying gives us the equation

\begin{equation}
    \begin{aligned}
        &\left( e^{- \lambda T_{i - 2}} - e^{- \lambda T_{i - 1}} \right)(e^{- (\lambda + \gamma)T_{i - 1}} - e^{- (\lambda + \gamma)T_{i}}) \\
        &\hspace{10pt} < (e^{- \lambda T_{i - 1}} - e^{- \lambda T_{i}})(e^{- (\lambda + \gamma)T_{i - 2}} - e^{- (\lambda + \gamma)T_{i - 1}})
    \end{aligned}
\end{equation}

Expanding both sides and canceling out
\({- e}^{- (2\lambda + \gamma)T_{i - 1}}\) from both sides gives us a
left hand side of

\begin{equation}
    \begin{aligned}
        &\left( e^{- \lambda T_{i - 1}} \right)\left( e^{- {\gamma T}_{i - 1}} \right)\left( e^{- \lambda T_{i - 2}} \right) \\
        &- \left( e^{- \lambda T_{i}} \right)\left( e^{- {\gamma T}_{i}} \right)\left( e^{- \lambda T_{i - 2}} \right) \\
        &+ \left( e^{- \lambda T_{i - 1}} \right)\left( e^{- {\gamma T}_{i}} \right)\left( e^{- \lambda T_{i}} \right)
    \end{aligned}
\end{equation}

which is lesser than the right hand side of

\begin{equation}
    \begin{aligned}
        &\left( e^{- \lambda T_{i - 1}} \right)\left( e^{- {\gamma T}_{i - 2}} \right)\left( e^{- \lambda T_{i - 2}} \right) \\
        &- \left( e^{- \lambda T_{i}} \right)\left( e^{- {\gamma T}_{i - 2}} \right)\left( e^{- \lambda T_{i - 2}} \right) \\
        &+ \left( e^{- \lambda T_{i - 1}} \right)\left( e^{- {\gamma T}_{i - 1}} \right)\left( e^{- \lambda T_{i}} \right)
    \end{aligned}
\end{equation}

We know that \(T_{i} > T_{i - 1} > T_{i - 2}\), so we can see that
\(e^{{- T}_{i - 2}} > e^{{- T}_{i - 1}} > e^{{- T}_{i}}\). Thus, it is
easy to see that the third term on the right hand side is larger than
the third term on the left hand side.

\begin{equation}
    \begin{aligned}
        &\left( e^{- \lambda T_{i - 1}} \right)\left( e^{- {\gamma T}_{i - 1}} \right)\left( e^{- \lambda T_{i}} \right) \\
        &\hspace{10pt} > \left( e^{- \lambda T_{i - 1}} \right)\left( e^{- {\gamma T}_{i}} \right)\left( e^{- \lambda T_{i}} \right)
    \end{aligned}
\end{equation}

Now, we must evaluate the remaining two terms together. The
\(\left( e^{- \lambda T_{i - 2}} \right)\) cancels out from both sides,
leaving us with

\begin{equation}
    \begin{aligned}
        &\left( e^{- \lambda T_{i - 1}} \right)\left( e^{- {\gamma T}_{i - 1}} \right) - \left( e^{- \lambda T_{i}} \right)\left( e^{- {\gamma T}_{i}} \right) \\
        &\hspace{10pt} < \left( e^{- \lambda T_{i - 1}} \right)\left( e^{- {\gamma T}_{i - 2}} \right) - \left( e^{- \lambda T_{i}} \right)\left( e^{- {\gamma T}_{i - 2}} \right)
    \end{aligned}
\end{equation}

If we factor out
\(\left( e^{- \lambda T_{i - 1}} \right)\left( e^{- {\gamma T}_{i - 1}} \right)\)
from the left hand side and
\(\left( e^{- \lambda T_{i - 1}} \right)\left( e^{- {\gamma T}_{i - 2}} \right)\)
from the right hand side, we get

\begin{equation}
    \begin{aligned}
        &\left( e^{- \lambda T_{i - 1}} \right)\left( e^{- {\gamma T}_{i - 1}} \right)\left( 1 - e^{(\lambda + \gamma)\left( T_{i - 1}{- T}_{i} \right)} \right) \\
        &\hspace{10pt} < \left( e^{- \lambda T_{i - 1}} \right)\left( e^{- {\gamma T}_{i - 2}} \right)(1 - e^{(\lambda)\left( T_{i - 1}{- T}_{i} \right)})
    \end{aligned}
\end{equation}

If we cancel like terms and factor out
\(e^{(\gamma)\left( T_{i - 1}{- T}_{i} \right)}\) from the left hand
side, we are left with

\begin{equation}
    \begin{aligned}
        &e^{- (\gamma)\left( T_{i} \right)}\left( \frac{1}{e^{(\gamma)\left( T_{i - 1}{- T}_{i} \right)}} - e^{(\lambda)\left( T_{i - 1}{- T}_{i} \right)} \right) \\
        &\hspace{10pt} < \left( e^{- {\gamma T}_{i - 2}} \right)\left(1 - e^{(\lambda)\left( T_{i - 1}{- T}_{i} \right)}\right)
    \end{aligned}
\end{equation}

Analyzing this equation, we see that we can rearrange this equation to

\begin{equation}
    e^{(\gamma)\left( {T_{i - 2} - T}_{i} \right)} < \frac{(1 - e^{(\lambda)\left( T_{i - 1}{- T}_{i} \right)})}{\left( \frac{1}{e^{(\gamma)\left( T_{i - 1}{- T}_{i} \right)}} - e^{(\lambda)\left( T_{i - 1}{- T}_{i} \right)} \right)}
\end{equation}

We recognize that if
\(e^{(\lambda)\left( T_{i - 1}{- T}_{i} \right)} = 0\), we would have
the equation

\begin{equation}
    e^{(\gamma)\left( {T_{i - 2} - T}_{i} \right)} < e^{(\gamma)\left( T_{i - 1}{- T}_{i} \right)}
\end{equation}

We see that if we define

\begin{equation}
    d = e^{T_{i - 1}{- T}_{i}}
\end{equation}

We can simplify the equation to

\begin{align}
    e^{(\gamma)\left( {T_{i - 2} - T}_{i} \right)} &< \frac{(1 - d^{(\lambda)})}{\left( \frac{1}{d^{(\gamma)}} - d^{(\lambda)} \right)} \\
    e^{(\gamma)\left( {T_{i - 2} - T}_{i} \right)} &< \frac{(d^{(\gamma)} - {d^{(\gamma)}d}^{(\lambda)})}{\left( 1 - {d^{(\gamma)}d}^{(\lambda)} \right)}
\end{align}

We know that \(0\  < \ d\  < \ 1\), so we check the limits on both
sides.

\begin{align}
    \lim_{d \rightarrow 0}\frac{(d^{(\gamma)} - {d^{(\gamma)}d}^{(\lambda)})}{\left( 1 - {d^{(\gamma)}d}^{(\lambda)} \right)} &= 0 \\
    \lim_{d \rightarrow 1}\frac{(d^{(\gamma)} - {d^{(\gamma)}d}^{(\lambda)})}{\left( 1 - {d^{(\gamma)}d}^{(\lambda)} \right)} &= \frac{0}{0}
\end{align}

We use L'hopital's Rule on the limit as d approaches 1 to get

\begin{equation}
    \lim_{d \rightarrow 1}\frac{({\gamma d}^{(\gamma - 1)} - {(\lambda + \gamma)d}^{(\lambda + \gamma - 1)})}{- {(\lambda + \gamma)d}^{(\lambda + \gamma - 1)}} = \frac{\lambda}{\gamma + \lambda}
\end{equation}

We see that
\(\frac{(d^{(\gamma)} - {d^{(\gamma)}d}^{(\lambda)})}{\left( 1 - {d^{(\gamma)}d}^{(\lambda)} \right)}\)
is strictly increasing with $\gamma$ because the \(d^{(\gamma)}\) term gets
larger as $\gamma$ gets larger, meaning the numerator increases faster than the
denominator. Thus, we see that
\(e^{(\gamma)\left( {T_{i - 2} - T}_{i} \right)}\) has an upper bound of
\(\frac{\lambda}{\gamma + \lambda}\) and a lower bound of 0. This means
the minimum time step from \(T_{i - 2}\) to \(T_{i}\) must be greater
than or equal to
\(\left| \frac{1}{\gamma}\ln\frac{\lambda}{\gamma + \lambda} \right|\)
for \(\tau_{i - 1}^{*} > \tau_{i}^{*}\). Otherwise, we have
\(\tau_{i - 1}^{*} \leq \tau_{i}^{*}\).

\end{appendixproof}

\begin{appendixproof}{COROLLARY~\ref{cor6}}
We wish to prove that, in the general case with $n$ stages of degradation
with \(T_{0} = 0\) and \(T_{n} = \infty\),
\(\tau_{i - 1}^{*} > \tau_{i}^{*}\). We begin this proof by plugging in
\(\tau_{i}^{*} = \sqrt{\frac{b + a\left( 1 - \frac{{2y}_{i}}{Z_{i}} \right)}{ab}}\).

\begin{equation}
    \sqrt{\frac{b + a\left( 1 - \frac{{2y}_{i - 1}}{Z_{i - 1}} \right)}{ab}} > \sqrt{\frac{b + a\left( 1 - \frac{{2y}_{i}}{Z_{i}} \right)}{ab}}
\end{equation}

Cancelling all the similarities by squaring both sides, multiplying by
$ab$, subtracting $b$, dividing by $a$, and subtracting 1
from both sides leaves us with

\begin{equation}
    - \frac{{2y}_{i - 1}}{Z_{i - 1}} > - \frac{{2y}_{i}}{Z_{i}}
\end{equation}

Now, we divide by -2 and reverse the comparison to reach the equation

\begin{equation}
    \frac{y_{i - 1}}{Z_{i - 1}} < \frac{y_{i}}{Z_{i}}
\end{equation}

We need to find the conditions under which this equation is true.

First, we substitute in the equations for
\(y_{i - 1},\ y_{i},\ Z_{i - 1},Z_{i}\ \)to get

\begin{equation}
    \frac{\int_{T_{i - 2}}^{T_{i - 1}}{e^{- \lambda t}dt}}{\int_{T_{i - 2}}^{T_{i - 1}}{e^{- \lambda t}e^{\gamma t}dt}} < \frac{\int_{T_{i - 1}}^{T_{i}}{e^{- \lambda t}dt}}{\int_{T_{i - 1}}^{T_{i}}{e^{- \lambda t}e^{\gamma t}dt}}
\end{equation}

Integrating all parts gives us the equation

\begin{equation}
    \begin{aligned}
        &\frac{\frac{1}{\lambda}(e^{- \lambda T_{i - 2}} - e^{- \lambda T_{i - 1}})}{\frac{1}{\lambda - \gamma}(e^{- (\lambda - \gamma)T_{i - 2}} - e^{- (\lambda - \gamma)T_{i - 1}})} \\
        &\hspace{10pt} < \frac{\frac{1}{\lambda}(e^{- \lambda T_{i - 1}} - e^{- \lambda T_{i}})}{\frac{1}{\lambda - \gamma}(e^{- (\lambda - \gamma)T_{i - 1}} - e^{- (\lambda - \gamma)T_{i}})}
    \end{aligned}
\end{equation}

Canceling out \(\frac{1}{\lambda}\) and \(\frac{1}{\lambda - \gamma}\)
and cross multiplying gives us the equation

\begin{equation}
    \begin{aligned}
        &\left( e^{- \lambda T_{i - 2}} - e^{- \lambda T_{i - 1}} \right)(e^{- (\lambda - \gamma)T_{i - 1}} - e^{- (\lambda - \gamma)T_{i}}) \\
        &\hspace{10pt} < (e^{- \lambda T_{i - 1}} - e^{- \lambda T_{i}})(e^{- (\lambda - \gamma)T_{i - 2}} - e^{- (\lambda - \gamma)T_{i - 1}})
    \end{aligned}
\end{equation}

Expanding both sides and canceling out
\({- e}^{- (2\lambda - \gamma)T_{i - 1}}\) from both sides gives us a
left hand side of

\begin{equation}
    \begin{aligned}
        &\left( e^{- \lambda T_{i - 1}} \right)\left( e^{{\gamma T}_{i - 1}} \right)\left( e^{- \lambda T_{i - 2}} \right) \\ 
        &- \left( e^{- \lambda T_{i}} \right)\left( e^{{\gamma T}_{i}} \right)\left( e^{- \lambda T_{i - 2}} \right) \\
        &+ \left( e^{- \lambda T_{i - 1}} \right)\left( e^{{\gamma T}_{i}} \right)\left( e^{- \lambda T_{i}} \right)
    \end{aligned}
\end{equation}

which is lesser than the right hand side of

\begin{equation}
    \begin{aligned}
        &\left( e^{- \lambda T_{i - 1}} \right)\left( e^{{\gamma T}_{i - 2}} \right)\left( e^{- \lambda T_{i - 2}} \right) \\
        &- \left( e^{- \lambda T_{i}} \right)\left( e^{{\gamma T}_{i - 2}} \right)\left( e^{- \lambda T_{i - 2}} \right) \\
        &+ \left( e^{- \lambda T_{i - 1}} \right)\left( e^{{\gamma T}_{i - 1}} \right)\left( e^{- \lambda T_{i}} \right)
    \end{aligned}
\end{equation}

We know that \(T_{i} > T_{i - 1} > T_{i - 2}\), so we can see that
\(e^{{- T}_{i - 2}} > e^{{- T}_{i - 1}} > e^{{- T}_{i}}\ \)and
\(e^{T_{i - 2}} < e^{T_{i - 1}} < e^{T_{i}}.\) Thus, it is easy to see
that the third term on the right hand side is lower than the third term
on the left hand side.

\begin{equation}
    \begin{aligned}
        &\left( e^{- \lambda T_{i - 1}} \right)\left( e^{{\gamma T}_{i - 1}} \right)\left( e^{- \lambda T_{i}} \right) \\
        &\hspace{10pt} < \left( e^{- \lambda T_{i - 1}} \right)\left( e^{{\gamma T}_{i}} \right)\left( e^{- \lambda T_{i}} \right)
    \end{aligned}
\end{equation}

Now, we must evaluate the remaining two terms together. The
\(\left( e^{- \lambda T_{i - 2}} \right)\) cancels out from both sides,
leaving us with

\begin{equation}
    \begin{aligned}
        &\left( e^{- \lambda T_{i - 1}} \right)\left( e^{{\gamma T}_{i - 1}} \right) - \left( e^{- \lambda T_{i}} \right)\left( e^{{\gamma T}_{i}} \right) \\
        &\hspace{10pt} < \left( e^{- \lambda T_{i - 1}} \right)\left( e^{{\gamma T}_{i - 2}} \right) - \left( e^{- \lambda T_{i}} \right)\left( e^{{\gamma T}_{i - 2}} \right)
    \end{aligned}
\end{equation}

If we factor out
\(\left( e^{- \lambda T_{i - 1}} \right)\left( e^{{\gamma T}_{i - 1}} \right)\)
from the left hand side and
\(\left( e^{- \lambda T_{i - 1}} \right)\left( e^{{\gamma T}_{i - 2}} \right)\)
from the right hand side, we get

\begin{equation}
    \begin{aligned}
        &\left( e^{- \lambda T_{i - 1}} \right)\left( e^{{\gamma T}_{i - 1}} \right)\left( 1 - e^{(\lambda - \gamma)\left( T_{i - 1}{- T}_{i} \right)} \right) \\
        &\hspace{10pt} < \left( e^{- \lambda T_{i - 1}} \right)\left( e^{{\gamma T}_{i - 2}} \right)(1 - e^{(\lambda)\left( T_{i - 1}{- T}_{i} \right)})
    \end{aligned}
\end{equation}

If we cancel like terms and factor out
\(e^{( - \gamma)\left( T_{i - 1}{- T}_{i} \right)}\) from the left hand
side, we are left with

\begin{equation}
    \begin{aligned}
        &e^{(\gamma)\left( T_{i} \right)}\left( \frac{1}{e^{( - \gamma)\left( T_{i - 1}{- T}_{i} \right)}} - e^{(\lambda)\left( T_{i - 1}{- T}_{i} \right)} \right) \\
        &\hspace{10pt} < \left( e^{{\gamma T}_{i - 2}} \right)\left(1 - e^{(\lambda)\left( T_{i - 1}{- T}_{i} \right)}\right)
    \end{aligned}
\end{equation}

Analyzing this equation, we see that we can rearrange this equation to

\begin{equation}
    e^{(\gamma)\left( {T_{i} - T}_{i - 2} \right)} < \frac{(1 - e^{(\lambda)\left( T_{i - 1}{- T}_{i} \right)})}{\left( \frac{1}{e^{( - \gamma)\left( T_{i - 1}{- T}_{i} \right)}} - e^{(\lambda)\left( T_{i - 1}{- T}_{i} \right)} \right)}
\end{equation}

We recognize that if
\(e^{(\lambda)\left( T_{i - 1}{- T}_{i} \right)} = 0\), we would have
the equation

\begin{equation}
    e^{(\gamma)\left( {T_{i} - T}_{i - 2} \right)} < e^{( - \gamma)\left( T_{i - 1}{- T}_{i} \right)}
\end{equation}

We see that if we define

\begin{equation}
    d = e^{T_{i - 1}{- T}_{i}}
\end{equation}

We can simplify the equation to

\begin{align}
    e^{(\gamma)\left( {T_{i - 2} - T}_{i} \right)} &< \frac{(1 - d^{(\lambda)})}{\left( \frac{1}{d^{(-\gamma)}} - d^{(\lambda)} \right)} \\
    e^{(-\gamma)\left( {T_{i - 2} - T}_{i} \right)} &< \frac{(d^{(-\gamma)} - {d^{(-\gamma)}d}^{(\lambda)})}{\left( 1 - {d^{(-\gamma)}d}^{(\lambda)} \right)}
\end{align}

We know that \(0\  < \ d\  < \ 1\), so we check the limits on both
sides.

\begin{align}
    \lim_{d \rightarrow 0}\frac{(d^{(-\gamma)} - {d^{(-\gamma)}d}^{(\lambda)})}{\left( 1 - {d^{(-\gamma)}d}^{(\lambda)} \right)} &= 0 \\
    \lim_{d \rightarrow 1}\frac{(d^{(-\gamma)} - {d^{(-\gamma)}d}^{(\lambda)})}{\left( 1 - {d^{(-\gamma)}d}^{(\lambda)} \right)} &= \frac{0}{0}
\end{align}

We use L'hopital's Rule on the limit as d approaches 1 to get

\begin{equation}
    \lim_{d \rightarrow 1}\frac{({-\gamma d}^{(-\gamma - 1)} - {(\lambda - \gamma)d}^{(\lambda - \gamma - 1)})}{- {(\lambda - \gamma)d}^{(\lambda - \gamma - 1)}} = \frac{\lambda}{\lambda-\gamma}
\end{equation}

We see that
\(\frac{(d^{(-\gamma)} - {d^{(-\gamma)}d}^{(\lambda)})}{\left( 1 - {d^{(-\gamma)}d}^{(\lambda)} \right)}\)
is decreasing with $\gamma$ because the \(d^{(-\gamma)}\) term gets
smaller as $\gamma$ gets larger, meaning the numerator decreases faster than the
denominator. Thus, we see that
\(e^{(\gamma)\left( {T_{i} - T}_{i - 2} \right)}\) has an upper bound of
\(\frac{\lambda}{\lambda - \gamma}\) and a lower bound of 0. This means
the minimum time step from \(T_{i - 2}\) to \(T_{i}\) must be lower than
or equal to
\(\left| \frac{1}{\gamma}\ln\frac{\lambda - \gamma}{\lambda} \right|\), with $0<\gamma<\lambda$
for \(\tau_{i - 1}^{*} > \tau_{i}^{*}\). Otherwise, we have
\(\tau_{i - 1}^{*} \leq \tau_{i}^{*}\).

\end{appendixproof}

\begin{appendixproof}{COROLLARY~\ref{cor3}}
    The parameters \(a\) and \(b\) will have a direct effect on
\(T_{i - 1}^{*}\) through the probability functions \(p\) and \(q\), as
they are found in the equations for \(p\) and \(q\) and thus have an
immediate impact on \(T_{i - 1}^{*}\). However, parameters \(a\) and
\(b\) have an indirect effect on \(T_{i - 1}^{*}\)
through\({\ \tau}_{i}^{*}\), \(y_{i}\) and \(Z_{i}\) because while \(a\)
and \(b\) are located in the equations for \({\ \tau}_{i}^{*}\),
\(y_{i}\) and \(Z_{i}\), none of these are located directly in equation
for \(T_{i - 1}^{*}\) and thus only exert an indirect effect. By the
envelope theorem, we can ignore the indirect effects on
\(T_{i - 1}^{*}\), allowing us to ignore terms like
\(\frac{d\tau_{i}^{*}\ }{da}\). To calculate the direct effect, recall

\begin{equation}
    p(\tau) = 1 - \frac{a}{2}\tau^{2} \ \text{ and } \ q(\tau) = 1 - \frac{b}{2}\tau^{2}
\end{equation}

Plug these into \(T_{i - 1}^{*}\):

\begin{equation}
\begin{aligned}
    &\hspace{98pt} T_{i - 1}^{*} = \frac{1}{\gamma}\ln \\
    &\left\lbrack \frac{\left(1 - \frac{a}{2}{\tau_{i - 1}^{*}}^{2}\right)\left(1 - \frac{b}{2}{\tau_{i - 1}^{*}}^{2}\right) - \left(1 - \frac{a}{2}{\tau_{i}^{*}}^{2}\right)\left(1 - \frac{b}{2}{\tau_{i}^{*}}^{2}\right)}{2\left( \left(1 - \frac{a}{2}{\tau_{i - 1}^{*}}^{2}\right) - \left(1 - \frac{a}{2}{\tau_{i}^{*}}^{2}\right) \right)} \right\rbrack 
\end{aligned}
\end{equation}

This equation simplifies to

\begin{equation}
    T_{i - 1}^{*} = \frac{1}{\gamma}\ln
\end{equation}

\begin{equation}
    \left\lbrack \frac{\left(1 - \frac{a}{2}{\tau_{i - 1}^{*}}^{2}\right)\left(1 - \frac{b}{2}{\tau_{i - 1}^{*}}^{2}\right) - \left(1 - \frac{a}{2}{\tau_{i}^{*}}^{2}\right)\left(1 - \frac{b}{2}{\tau_{i}^{*}}^{2}\right)}{a\left( {\tau_{i}^{*}}^{2} - {\tau_{i - 1}^{*}}^{2} \right)} \right\rbrack
\end{equation}

Let

\begin{equation}
    \begin{aligned}
        f(a,b) &= \left(1 - \frac{a}{2}{\tau_{i - 1}^{*}}^{2}\right)\left(1 - \frac{b}{2}{\tau_{i - 1}^{*}}^{2}\right) \\
        &\hspace{10pt} - \left(1 - \frac{a}{2}{\tau_{i}^{*}}^{2}\right)\left(1 - \frac{b}{2}{\tau_{i}^{*}}^{2}\right)
    \end{aligned}
\end{equation}

\begin{align}
    f^{\prime}(a,b) &= \frac{\partial f(a,b)}{\partial a} \nonumber \\
    &= \left(\frac{1}{2}{\tau_{i}^{*}}^{2}\right)\left( 1 - \frac{b}{2}{\tau_{i}^{*}}^{2} \right) \\
    &\hspace{10pt}- \left(\frac{1}{2}{\tau_{i - 1}^{*}}^{2}\right)\left(1 - \frac{b}{2}{\tau_{i - 1}^{*}}^{2}\right) \nonumber
\end{align}

Since the denominator of \(T^{*}\) is negative with
\(a({\tau_{i}^{*}}^{2} - {\tau_{i - 1}^{*}}^{2}) < 0\), we know that
\(f(a,b) < 0\), otherwise the \(T^{*}\) equation would be invalid.

To find the conditions for which
\(\frac{{\partial T}^{*}}{\partial a} > 0\), we will differentiate
\(T^{*}\) with respect to $a$

\begin{equation}
    \frac{{\partial T}^{*}}{\partial a} = \frac{1}{\gamma}\frac{\frac{\partial}{\partial a}\left(\frac{f(a,b)}{a\left({\tau_{i}^{*}}^{2} - {\tau_{i - 1}^{*}}^{2}\right)}\right)}{\frac{f(a,b)}{a\left({\tau_{i}^{*}}^{2} - {\tau_{i - 1}^{*}}^{2}\right)}}
\end{equation}

Expanding the numerator gives us

\begin{equation}
    \frac{{\partial T}^{*}}{\partial a} = \frac{1}{\gamma}\frac{\frac{a\left({\tau_{i}^{*}}^{2} - {\tau_{i - 1}^{*}}^{2}\right)f^{\prime}(a,b) - \left({\tau_{i}^{*}}^{2} - {\tau_{i - 1}^{*}}^{2}\right)f(a,b)}{{\left(a\left({\tau_{i}^{*}}^{2} - {\tau_{i - 1}^{*}}^{2}\right)\right)}^{2}}}{\frac{f(a,b)}{a\left({\tau_{i}^{*}}^{2} - {\tau_{i - 1}^{*}}^{2}\right)}}
\end{equation}

We cancel the \(a({\tau_{i}^{*}}^{2} - {\tau_{i - 1}^{*}}^{2})\) term
and \(({\tau_{i}^{*}}^{2} - {\tau_{i - 1}^{*}}^{2})\) term from the
numerator and denominator to get

\begin{equation}
    \frac{{\partial T}^{*}}{\partial a} = \frac{1}{\gamma}\frac{af^{\prime}(a,b) - f(a,b)}{af(a,b)}
\end{equation}

We see that, for \(\frac{{\partial T}^{*}}{\partial a} > 0\), we must
fulfill the conditions that \(af^{\prime}(a,b) - f(a,b) < 0\) given that
\(f(a,b) < 0\) and \(\frac{1}{\gamma} > 0\). Plugging in for
\(f^{\prime}(a,b)\) and \(f(a,b)\), and plugging in
\(q(\tau) = 1 - \frac{b}{2}\tau^{2}\ \) for readability, we see that

\begin{equation}
    \begin{aligned}
        &\left( \frac{a}{2}{\tau_{i}^{*}}^{2} \right)q\left( \tau_{i}^{*} \right) - \left( \frac{a}{2}{\tau_{i - 1}^{*}}^{2} \right)q\left( \tau_{i - 1}^{*} \right) \\
        &- \left( 1 - \frac{a}{2}{\tau_{i - 1}^{*}}^{2} \right)q\left( \tau_{i - 1}^{*} \right) + \left( 1 - \frac{a}{2}{\tau_{i}^{*}}^{2} \right)q\left( \tau_{i}^{*} \right) < 0,
    \end{aligned}
\end{equation}

which simplifies to

\begin{equation}
    q\left( \tau_{i}^{*} \right) < q\left( \tau_{i - 1}^{*} \right),
\end{equation}

which is never true since \(\tau_{i - 1}^{*} > \tau_{i}^{*}\) by
\Cref{cor2} and \(q\) is always decreasing, meaning that
\(\frac{{\partial T}^{*}}{\partial a} < 0\) is always true. With the
condition of \(\frac{{\partial T}^{*}}{\partial a} < 0\) established, we
seek to establish the conditions for
\(\frac{{\partial T}^{*}}{\partial b} > 0\). First, we establish that

\begin{equation}
    \frac{{\partial T}^{*}}{\partial b} = \frac{1}{\gamma}\frac{\frac{\partial}{\partial b}\left(\frac{f(a,b)}{a\left({\tau_{i}^{*}}^{2} - {\tau_{i - 1}^{*}}^{2}\right)}\right)}{\frac{f(a,b)}{a\left({\tau_{i}^{*}}^{2} - {\tau_{i - 1}^{*}}^{2}\right)}} = \frac{1}{\gamma}\frac{\frac{\partial f(a,b)}{\partial b}}{f(a,b)}
\end{equation}

We now expand \(\frac{\partial f(ab)}{\partial b}\).

\begin{equation}
    \begin{aligned}
        \frac{\partial f(a,b)}{\partial b} &= \left( - \frac{{\tau_{i - 1}^{*}}^{2}}{2} \right)\left( 1 - \frac{a}{2}{\tau_{i - 1}^{*}}^{2} \right) \\
        &\hspace{10pt}+ \left(\frac{{\tau_{i}^{*}}^{2}}{2}\right)\left(1 - \frac{a}{2}{\tau_{i}^{*}}^{2}\right)
    \end{aligned}
\end{equation}

Since \(f(a,b) < 0,\) and \(\frac{1}{\gamma} > 0\), we see that the
condition for \(\frac{{\partial T}^{*}}{\partial b} > 0\) is

\begin{align}
    \frac{\partial f(a,b)}{\partial b} &< 0 \\
    \left( \frac{{\tau_{i - 1}^{*}}^{2}}{2} \right)\left( 1 - \frac{a}{2}{\tau_{i - 1}^{*}}^{2} \right) &> \left(\frac{{\tau_{i}^{*}}^{2}}{2}\right)\left(1 - \frac{a}{2}{\tau_{i}^{*}}^{2}\right)
\end{align}

Expanding and plugging in
\(\tau_{i}^{*} = \sqrt{\frac{b + a\left( 1 - \frac{{2y}_{i}}{Z_{i}} \right)}{ab}}\),
we get

\begin{equation}
    \begin{aligned}
        &\frac{b + a - \frac{{2ay}_{i - 1}}{Z_{i - 1}}}{ab} - \frac{a}{2}\left( \frac{b + a - \frac{{2ay}_{i - 1}}{Z_{i - 1}}}{ab} \right)^{2} \\
        &\hspace{10pt} > \frac{b + a - \frac{{2ay}_{i}}{Z_{i}}}{ab} - \frac{a}{2}{\left(\frac{b + a - \frac{{2ay}_{i}}{Z_{i}}}{ab}\right)}^{2}
    \end{aligned}
\end{equation}

Subtracting \(\frac{b + a}{ab}\) from both sides and multiplying both
sides by \(ab\) gives us

\begin{equation}
    \begin{aligned}
        &- \frac{{2ay}_{i - 1}}{Z_{i - 1}} - \frac{1}{2b}\left( b + a - \frac{{2ay}_{i - 1}}{Z_{i - 1}} \right)^{2} \\
        &\hspace{10pt} > - \frac{{2ay}_{i}}{Z_{i}} - \frac{1}{2b}{\left(b + a - \frac{{2ay}_{i}}{Z_{i}}\right)}^{2}
    \end{aligned}
\end{equation}

Expanding the quadratic on both sides, we get

\begin{equation}
    \begin{aligned}
        &- \frac{{2ay}_{i - 1}}{Z_{i - 1}} - \frac{1}{2b}\left( b^{2} + a^{2} + 2ab + \right. \\
        &\hspace{6pt}\left. \frac{4a^{2}y_{i - 1}^{2}}{Z_{i - 1}^{2}} - \frac{{4a^{2}y}_{i - 1}}{Z_{i - 1}} - \frac{{4aby}_{i - 1}}{Z_{i - 1}} \right) \\
        &\hspace{10pt} > - \frac{{2ay}_{i}}{Z_{i}} - \frac{1}{2b}\left( b^{2} + a^{2} + 2ab + \right. \\
        &\hspace{30pt}\left. \frac{4a^{2}y_{i}^{2}}{Z_{i}^{2}} - \frac{{4a^{2}y}_{i}}{Z_{i}} - \frac{{4aby}_{i}}{Z_{i}} \right)
    \end{aligned}
\end{equation}

Subtracting all the common terms, and we are left with

\begin{equation}
    \begin{aligned}
        &- \frac{{2ay}_{i - 1}}{Z_{i - 1}} - \frac{2a^{2}y_{i - 1}^{2}}{{bZ}_{i - 1}^{2}} + \frac{{2a^{2}y}_{i - 1}}{{bZ}_{i - 1}} + \frac{{2ay}_{i - 1}}{Z_{i - 1}} \\
        &\hspace{10pt} > - \frac{{2ay}_{i}}{Z_{i}} - \frac{2a^{2}y_{i}^{2}}{{bZ}_{i}^{2}} + \frac{{2a^{2}y}_{i}}{{bZ}_{i}} + \frac{{2ay}_{i}}{Z_{i}} \\
    \end{aligned}
\end{equation}

\begin{equation}
    - \frac{2a^{2}y_{i - 1}^{2}}{{bZ}_{i - 1}^{2}} + \frac{{2a^{2}y}_{i - 1}}{{bZ}_{i - 1}} > - \frac{2a^{2}y_{i}^{2}}{{bZ}_{i}^{2}} + \frac{{2a^{2}y}_{i}}{{bZ}_{i}}
\end{equation}

\begin{equation}
    \frac{y_{i - 1}}{Z_{i - 1}} - \frac{y_{i}}{Z_{i}} > \frac{y_{i - 1}^{2}}{Z_{i - 1}^{2}} - \frac{y_{i}^{2}}{Z_{i}^{2}}
\end{equation}

\begin{equation}
    \frac{y_{i - 1}}{Z_{i - 1}} - \frac{y_{i}}{Z_{i}} > \left( \frac{y_{i - 1}}{Z_{i - 1}} - \frac{y_{i}}{Z_{i}} \right) \cdot \left( \frac{y_{i - 1}}{Z_{i - 1}} + \frac{y_{i}}{Z_{i}} \right)
\end{equation}

By \Cref{cor2}, we see that
\(\frac{y_{i - 1}}{Z_{i - 1}} - \frac{y_{i}}{Z_{i}} < 0\) and we know
that that \(y_{i} > Z_{i}\). Dividing both sides by
\(\frac{y_{i - 1}}{Z_{i - 1}} - \frac{y_{i}}{Z_{i}}\) flips the
inequality sign to give us

\begin{equation}
    \frac{y_{i - 1}}{Z_{i - 1}} + \frac{y_{i}}{Z_{i}} > 1
\end{equation}

This condition is always true because \(y_{i} > Z_{i}\), meaning
\(\frac{{\partial T}^{*}}{\partial b} > 0\) is always true.

\end{appendixproof}

\begin{appendixproof}{COROLLARY~\ref{cor7}}

\begin{equation}
    p(\tau) = 1 - \frac{a}{2}\tau^{2}\ \text{ and }\ q(\tau) = 1 - \frac{b}{2}\tau^{2}
\end{equation}

Plug these into \(T_{i - 1}^{*}\):

\begin{equation}
    T_{i - 1}^{*} = \frac{1}{\gamma}\ln
\end{equation}

\begin{equation}
    \left\lbrack \frac{2\left( \left(1 - \frac{a}{2}{\tau_{i - 1}^{*}}^{2}\right) - \left(1 - \frac{a}{2}{\tau_{i}^{*}}^{2}\right) \right)}
    {\left(1 - \frac{a}{2}{\tau_{i - 1}^{*}}^{2}\right)\left(1 - \frac{b}{2}{\tau_{i - 1}^{*}}^{2}\right) \left(1 - \frac{a}{2}{\tau_{i}^{*}}^{2}\right)\left(1 - \frac{b}{2}{\tau_{i}^{*}}^{2}\right)}
    \right\rbrack
\end{equation}

This equation simplifies to

\begin{equation}
    T_{i - 1}^{*} = \frac{1}{\gamma}\ln
\end{equation}

\begin{equation}
    \left\lbrack \frac{a\left( {\tau_{i}^{*}}^{2} - {\tau_{i - 1}^{*}}^{2} \right)}{\left(1 - \frac{a}{2}{\tau_{i - 1}^{*}}^{2}\right)\left(1 - \frac{b}{2}{\tau_{i - 1}^{*}}^{2}\right) - \left(1 - \frac{a}{2}{\tau_{i}^{*}}^{2}\right)\left(1 - \frac{b}{2}{\tau_{i}^{*}}^{2}\right)} \right\rbrack
\end{equation}
\[\]

Let

\begin{align}
    f(a,b) &= \left(1 - \frac{a}{2}{\tau_{i - 1}^{*}}^{2}\right)\left(1 - \frac{b}{2}{\tau_{i - 1}^{*}}^{2}\right) \nonumber \\
    &\hspace{10pt}- \left(1 - \frac{a}{2}{\tau_{i}^{*}}^{2}\right)\left(1 - \frac{b}{2}{\tau_{i}^{*}}^{2}\right) \\
    f^{\prime}(a,b) &= \frac{\partial f(a,b)}{\partial a} \nonumber \\
    &= \left(\frac{1}{2}{\tau_{i}^{*}}^{2}\right)\left( 1 - \frac{b}{2}{\tau_{i}^{*}}^{2} \right) \nonumber \\
    &\hspace{10pt}- \left(\frac{1}{2}{\tau_{i - 1}^{*}}^{2}\right)\left(1 - \frac{b}{2}{\tau_{i - 1}^{*}}^{2}\right)
\end{align}

Since \({\tau_{i - 1}^{*}} > {\tau_{i}^{*}}\) when $0<\gamma< \lambda$, the numerator of \(T^{*}\) is negative with
\(a({\tau_{i}^{*}}^{2} - {\tau_{i - 1}^{*}}^{2}) < 0\), we know that
\(f(a,b) < 0\), otherwise the \(T^{*}\) equation would be invalid.

To find the conditions for which
\(\frac{{\partial T}^{*}}{\partial a} > 0\), we will differentiate
\(T^{*}\) with respect to $a$

\begin{equation}
    \frac{{\partial T}^{*}}{\partial a} = \frac{1}{\gamma}\frac{\frac{\partial}{\partial a}\left(\frac{a\left({\tau_{i}^{*}}^{2} - {\tau_{i - 1}^{*}}^{2}\right)}{f(a,b)}\right)}{\frac{a\left({\tau_{i}^{*}}^{2} - {\tau_{i - 1}^{*}}^{2}\right)}{f(a,b)}}
\end{equation}

Expanding the numerator gives us

\begin{equation}
    \frac{{\partial T}^{*}}{\partial a} = \frac{1}{\gamma}\frac{\frac{\left( {\tau_{i}^{*}}^{2} - {\tau_{i - 1}^{*}}^{2} \right)f(a,b) - a({\tau_{i}^{*}}^{2} - {\tau_{i - 1}^{*}}^{2})f^{\prime}(a,b)}{{(f(a,b))}^{2}}}{\frac{a({\tau_{i}^{*}}^{2} - {\tau_{i - 1}^{*}}^{2})}{f(a,b)}}
\end{equation}

We cancel the \(f(a,b)\) term and
\(({\tau_{i}^{*}}^{2} - {\tau_{i - 1}^{*}}^{2})\) term from the
numerator and denominator to get

\begin{equation}
    \frac{{\partial T}^{*}}{\partial a} = \frac{1}{\gamma}\frac{f(a,b) - af^{\prime}(a,b)}{af(a,b)}
\end{equation}

We see that, for \(\frac{{\partial T}^{*}}{\partial a} > 0\), we must
fulfill the conditions that \(f(a,b) - af^{\prime}(a,b) < 0\) given that
\(f(a,b) < 0\) and \(\frac{1}{\gamma} > 0\). Plugging in for
\(f^{\prime}(a,b)\) and \(f(a,b)\), and plugging in
\(q(\tau) = 1 - \frac{b}{2}\tau^{2}\ \) for readability, we see that

\begin{equation}
    \begin{aligned}
        &\left( 1 - \frac{a}{2}{\tau_{i - 1}^{*}}^{2} \right)q\left( \tau_{i - 1}^{*} \right) - \left( 1 - \frac{a}{2}{\tau_{i}^{*}}^{2} \right)q\left( \tau_{i}^{*} \right) \\
        &- \left( \frac{a}{2}{\tau_{i}^{*}}^{2} \right)q\left( \tau_{i}^{*} \right) + \left( \frac{a}{2}{\tau_{i - 1}^{*}}^{2} \right)q\left( \tau_{i - 1}^{*} \right) < 0,
    \end{aligned}
\end{equation}

which simplifies to

\begin{equation}
    q\left( \tau_{i - 1}^{*} \right) < q\left( \tau_{i}^{*} \right),
\end{equation}

which is true since \(\tau_{i - 1}^{*} > \tau_{i}^{*}\) by
\cref{cor6} and \(q\) is decreasing, meaning that
\(\frac{{\partial T}^{*}}{\partial a} > 0\) is true. With the
condition of \(\frac{{\partial T}^{*}}{\partial a} > 0\) established, we
seek to establish the conditions for
\(\frac{{\partial T}^{*}}{\partial b} > 0\). First, we establish that

\begin{equation}
    \begin{aligned}
        \frac{{\partial T}^{*}}{\partial b} &= \frac{1}{\gamma}\frac{\frac{\partial}{\partial b}\left(\frac{a\left({\tau_{i}^{*}}^{2} - {\tau_{i - 1}^{*}}^{2}\right)}{f(a,b)}\right)}{\frac{a\left({\tau_{i}^{*}}^{2} - {\tau_{i - 1}^{*}}^{2}\right)}{f(a,b)}} \\
        &= \frac{1}{\gamma}f(a,b)\frac{\partial}{\partial b}\left( \frac{1}{f(a,b)} \right)
    \end{aligned}
\end{equation}

We now expand $\frac{\partial}{\partial b}\left( \frac{1}{f(a,b)} \right)$.

\begin{align}
    \frac{\partial}{\partial b}\left( \frac{1}{f(a,b)} \right) = \  - \frac{f^{\prime}(a,b)}{f(a,b)^{2}} \hspace{32pt} \\
    \frac{\left( \frac{{\tau_{i - 1}^{*}}^{2}}{2} \right)\left( 1 - \frac{a}{2}{\tau_{i - 1}^{*}}^{2} \right) - \left( \frac{{\tau_{i}^{*}}^{2}}{2} \right)\left( 1 - \frac{a}{2}{\tau_{i}^{*}}^{2} \right)}{f(a,b)^{2}}
\end{align}

Since \(f(a,b) < 0,\) and \(\frac{1}{\gamma} > 0\), we see that the
condition for \(\frac{{\partial T}^{*}}{\partial b} > 0\) is

\begin{equation}
    \left( \frac{{\tau_{i - 1}^{*}}^{2}}{2} \right)\left( 1 - \frac{a}{2}{\tau_{i - 1}^{*}}^{2} \right) - \left( \frac{{\tau_{i}^{*}}^{2}}{2} \right)\left( 1 - \frac{a}{2}{\tau_{i}^{*}}^{2} \right) < 0
\end{equation}

\begin{equation}
    \left( \frac{{\tau_{i - 1}^{*}}^{2}}{2} \right)\left( 1 - \frac{a}{2}{\tau_{i - 1}^{*}}^{2} \right) < \left(\frac{{\tau_{i}^{*}}^{2}}{2}\right)\left(1 - \frac{a}{2}{\tau_{i}^{*}}^{2}\right)
\end{equation}

Expanding and plugging in
\(\tau_{i}^{*} = \sqrt{\frac{b + a\left( 1 - \frac{{2y}_{i}}{Z_{i}} \right)}{ab}}\),
we get

\begin{equation}
    \begin{aligned}
        &\frac{b + a - \frac{{2ay}_{i - 1}}{Z_{i - 1}}}{ab} - \frac{a}{2}\left( \frac{b + a - \frac{{2ay}_{i - 1}}{Z_{i - 1}}}{ab} \right)^{2} \\
        &\hspace{10pt} < \frac{b + a - \frac{{2ay}_{i}}{Z_{i}}}{ab} - \frac{a}{2}{\left(\frac{b + a - \frac{{2ay}_{i}}{Z_{i}}}{ab}\right)}^{2}
    \end{aligned}
\end{equation}

Subtracting \(\frac{b + a}{ab}\) from both sides and multiplying both
sides by \(ab\) gives us

\begin{equation}
    \begin{aligned}
        &- \frac{{2ay}_{i - 1}}{Z_{i - 1}} - \frac{1}{2b}\left( b + a - \frac{{2ay}_{i - 1}}{Z_{i - 1}} \right)^{2} \\
        &\hspace{10pt} < - \frac{{2ay}_{i}}{Z_{i}} - \frac{1}{2b}{\left(b + a - \frac{{2ay}_{i}}{Z_{i}}\right)}^{2}
    \end{aligned}
\end{equation}

Expanding the quadratic on both sides, we get

\begin{equation}
    \begin{aligned}
        &- \frac{{2ay}_{i - 1}}{Z_{i - 1}} - \frac{1}{2b}\left( b^{2} + a^{2} + 2ab + \right. \\
        &\hspace{75pt} \left. \frac{4a^{2}y_{i - 1}^{2}}{Z_{i - 1}^{2}} - \frac{{4a^{2}y}_{i - 1}}{Z_{i - 1}} - \frac{{4aby}_{i - 1}}{Z_{i - 1}} \right) \\
        &\hspace{10pt} < - \frac{{2ay}_{i}}{Z_{i}} - \frac{1}{2b}\left( b^{2} + a^{2} + 2ab + \right. \\
        &\hspace{75pt} \left. \frac{4a^{2}y_{i}^{2}}{Z_{i}^{2}} - \frac{{4a^{2}y}_{i}}{Z_{i}} - \frac{{4aby}_{i}}{Z_{i}} \right)
    \end{aligned}
\end{equation}

Subtracting all the common terms, and we are left with

\begin{equation}
    \begin{aligned}
        &- \frac{{2ay}_{i - 1}}{Z_{i - 1}} - \frac{2a^{2}y_{i - 1}^{2}}{{bZ}_{i - 1}^{2}} + \frac{{2a^{2}y}_{i - 1}}{{bZ}_{i - 1}} + \frac{{2ay}_{i - 1}}{Z_{i - 1}} \\
        &\hspace{10pt} < - \frac{{2ay}_{i}}{Z_{i}} - \frac{2a^{2}y_{i}^{2}}{{bZ}_{i}^{2}} + \frac{{2a^{2}y}_{i}}{{bZ}_{i}} + \frac{{2ay}_{i}}{Z_{i}}
    \end{aligned}
\end{equation}

\begin{equation}
    - \frac{2a^{2}y_{i - 1}^{2}}{{bZ}_{i - 1}^{2}} + \frac{{2a^{2}y}_{i - 1}}{{bZ}_{i - 1}} < - \frac{2a^{2}y_{i}^{2}}{{bZ}_{i}^{2}} + \frac{{2a^{2}y}_{i}}{{bZ}_{i}}
\end{equation}

\begin{equation}
    \frac{y_{i - 1}}{Z_{i - 1}} - \frac{y_{i}}{Z_{i}} < \frac{y_{i - 1}^{2}}{Z_{i - 1}^{2}} - \frac{y_{i}^{2}}{Z_{i}^{2}}
\end{equation}

\begin{equation}
    \frac{y_{i - 1}}{Z_{i - 1}} - \frac{y_{i}}{Z_{i}} < \left( \frac{y_{i - 1}}{Z_{i - 1}} - \frac{y_{i}}{Z_{i}} \right) \cdot \left( \frac{y_{i - 1}}{Z_{i - 1}} + \frac{y_{i}}{Z_{i}} \right)
\end{equation}

By \Cref{cor6}, we see that
\(\frac{y_{i - 1}}{Z_{i - 1}} - \frac{y_{i}}{Z_{i}} < 0\) and we know
that that for all \(\gamma > 0,\ \ y_{i} < Z_{i}\). Dividing both sides
by \(\frac{y_{i - 1}}{Z_{i - 1}} - \frac{y_{i}}{Z_{i}}\) flips the
inequality sign to give us

\begin{equation}
    \frac{y_{i - 1}}{Z_{i - 1}} + \frac{y_{i}}{Z_{i}} < 1
\end{equation}

This condition is not always true unless \(\gamma \gg \lambda\), meaning
\(\frac{{\partial T}^{*}}{\partial b} < 0\) is generally true, unless
\(\gamma \gg \lambda\) and when \(\gamma > 0.\)

\end{appendixproof}

\begin{appendixproof}{COROLLARY~\ref{cor4}}

The conditions for
\(\frac{{\partial\tau}_{i}^{*}}{\partial\lambda} < 0\) can be calculated
by writing out their expansions. First, we analyze
\(\frac{{\partial\tau}_{i}^{*}}{\partial\lambda}\).

\begin{equation}
    \begin{aligned}
        \frac{{\partial\tau}_{i}^{*}}{\partial\lambda} &= \frac{\partial}{\partial\lambda}\left( \frac{b + a\left( 1 - \frac{{2y}_{i}}{Z_{i}} \right)}{ab} \right)^{0.5} \\
        &= \left( \frac{a}{ab} \right)^{0.5}\frac{\partial}{\partial\lambda}{\left(\frac{b}{a} + 1 - \frac{{2y}_{i}}{Z_{i}}\right)}^{0.5}
    \end{aligned}
\end{equation}

We assume that \(\tau_{i}^{*}\) is defined, so we will assume
\(\frac{b}{a} + 1 - \frac{{2y}_{i}}{Z_{i}} > 0\)

Since \(\left( \frac{a}{ab} \right)^{0.5}\ \)is always positive, the
conditions for \(\frac{{\partial\tau}_{i}^{*}}{\partial\lambda}\) can be
simplified down to

\begin{equation}
    \frac{{\partial\tau}_{i}^{*}}{\partial\lambda} = \frac{\partial}{\partial\lambda}{\left(\frac{b}{a} + 1 - \frac{{2y}_{i}}{Z_{i}}\right)}^{0.5} = \frac{2*\frac{\partial}{\partial\lambda}\left( - \frac{y_{i}}{Z_{i}}\right)}{2{\left(\frac{b}{a} + 1 - \frac{{2y}_{i}}{Z_{i}}\right)}^{0.5}}
\end{equation}

We note that the denominator is positive, meaning the direction of
\(\frac{{\partial\tau}_{i}^{*}}{\partial\lambda}\) can be determined by
the direction and conditions of equation

\begin{equation}
    \frac{\partial}{\partial\lambda}\left( - \frac{y_{i}}{Z_{i}}\right)
\end{equation}

Here, \(Z_{i}\) shrinks faster than \(y_{i}\) as \(\lambda\) increases.

This implies
\(\frac{\partial}{\partial\lambda}\left( \frac{y_{i}}{Z_{i}} \right) > 0\)

\(\frac{\partial}{\partial\lambda}\left( - \frac{y_{i}}{Z_{i}} \right)\)
\textless{} 0.

Therefore, \(\frac{{\partial\tau}_{i}^{*}}{\partial\lambda} < 0\)

In the same way, the conditions for
\(\frac{{\partial\tau}_{i}^{*}}{\partial\gamma} > 0\) can be calculated
by writing out their expansions. First, we analyze
\(\frac{{\partial\tau}_{i}^{*}}{\partial\gamma}\).

\begin{equation}
    \begin{aligned}
        \frac{{\partial\tau}_{i}^{*}}{\partial\gamma} &= \frac{\partial}{\partial\gamma}\left( \frac{b + a\left( 1 - \frac{{2y}_{i}}{Z_{i}} \right)}{ab} \right)^{0.5} \\
        &= \left( \frac{a}{ab} \right)^{0.5}\frac{\partial}{\partial\gamma}{\left(\frac{b}{a} + 1 - \frac{{2y}_{i}}{Z_{i}}\right)}^{0.5}
    \end{aligned}
\end{equation}

We assume that \(\tau_{i}^{*}\) is defined, so we will assume
\(\frac{b}{a} + 1 - \frac{{2y}_{i}}{Z_{i}} > 0\)

Since \(\left( \frac{a}{ab} \right)^{0.5}\ \)is always positive, the
conditions for \(\frac{{\partial\tau}_{i}^{*}}{\partial\gamma}\) can be
simplified down to

\begin{equation}
    \frac{{\partial\tau}_{i}^{*}}{\partial\gamma} = \frac{\partial}{\partial\gamma}{\left(\frac{b}{a} + 1 - \frac{{2y}_{i}}{Z_{i}}\right)}^{0.5} = \frac{2*\frac{\partial}{\partial\gamma}\left( - \frac{y_{i}}{Z_{i}}\right)}{2{\left(\frac{b}{a} + 1 - \frac{{2y}_{i}}{Z_{i}}\right)}^{0.5}}
\end{equation}

We note that the denominator is positive, meaning the direction of
\(\frac{{\partial\tau}_{i}^{*}}{\partial\gamma}\) can be determined by
the direction and conditions of equation

\begin{equation}
    \frac{\partial}{\partial\gamma}\left( - \frac{y_{i}}{Z_{i}}\right)
\end{equation}

Since \(y_{i}\) has no \(\gamma\) term, we are left evaluating equation

\begin{equation}
    {- y}_{i}\frac{\partial}{\partial\gamma}\left( \frac{1}{Z_{i}} \right) > 0
\end{equation}

\begin{equation}
    {- y}_{i}\frac{Z_{i}\frac{\partial}{\partial\gamma}(1) - (1)\frac{\partial}{\partial\gamma}(Z_{i})}{{(Z_{i})}^{2}} > 0
\end{equation}

\begin{equation}
    \frac{\partial}{\partial\gamma}\left( Z_{i} \right) > 0
\end{equation}

We know that
\({(\lambda + \gamma)Z}_{i} = e^{- (\lambda + \gamma)T_{i - 1}} - e^{- (\lambda + \gamma)T_{i}}\),
so we calculate that

\begin{equation}
    \frac{\partial}{\partial\gamma}\left( (\lambda + \gamma)Z_{i} \right) = T_{i}e^{- (\lambda + \gamma)T_{i}} - T_{i - 1}e^{- (\lambda + \gamma)T_{i - 1}}
\end{equation}

\begin{equation}
    Z_{i}\frac{\partial}{\partial\gamma}(\lambda + \gamma) + (\lambda + \gamma)\frac{\partial}{\partial\gamma}Z_{i} = T_{i}e^{- (\lambda + \gamma)T_{i}} - T_{i - 1}e^{- (\lambda + \gamma)T_{i - 1}}
\end{equation}

\begin{equation}
    \frac{dZ_{i}}{d\gamma} = \frac{T_{i}e^{- (\lambda + \gamma)T_{i}} - T_{i - 1}e^{- (\lambda + \gamma)T_{i - 1}} - Z_{i}}{(\lambda + \gamma)}
\end{equation}

From this, we can see \(\frac{dZ_{i}}{d\gamma} > 0\) is false. Hence
\(\frac{{\partial\tau}_{i}^{*}}{\partial\gamma} < 0.\)

\end{appendixproof}

\begin{appendixproof}{COROLLARY~\ref{cor8}}

The conditions for
\(\frac{{\partial\tau}_{i}^{*}}{\partial\lambda} < 0\) can be calculated
by writing out their expansions. First, we analyze
\(\frac{{\partial\tau}_{i}^{*}}{\partial\lambda}\).

\begin{equation}
    \begin{aligned}
        \frac{{\partial\tau}_{i}^{*}}{\partial\lambda} &= \frac{\partial}{\partial\lambda}\left( \frac{b + a\left( 1 - \frac{{2y}_{i}}{Z_{i}} \right)}{ab} \right)^{0.5} \\
        &= \left( \frac{a}{ab} \right)^{0.5}\frac{\partial}{\partial\lambda}{\left(\frac{b}{a} + 1 - \frac{{2y}_{i}}{Z_{i}}\right)}^{0.5}
    \end{aligned}
\end{equation}

We assume that \(\tau_{i}^{*}\) is defined, so we will assume
\(\frac{b}{a} + 1 - \frac{{2y}_{i}}{Z_{i}} > 0\)

Since \(\left( \frac{a}{ab} \right)^{0.5}\ \)is always positive, the
conditions for \(\frac{{\partial\tau}_{i}^{*}}{\partial\lambda}\) can be
simplified down to

\begin{equation}
    \frac{{\partial\tau}_{i}^{*}}{\partial\lambda} = \frac{\partial}{\partial\lambda}{\left(\frac{b}{a} + 1 - \frac{{2y}_{i}}{Z_{i}}\right)}^{0.5} = \frac{2*\frac{\partial}{\partial\lambda}\left( - \frac{y_{i}}{Z_{i}}\right)}{2{\left(\frac{b}{a} + 1 - \frac{{2y}_{i}}{Z_{i}}\right)}^{0.5}}
\end{equation}

We note that the denominator is positive, meaning the direction of
\(\frac{{\partial\tau}_{i}^{*}}{\partial\lambda}\) can be determined by
the direction and conditions of equation

\begin{equation}
    \frac{\partial}{\partial\lambda}\left( - \frac{y_{i}}{Z_{i}}\right)
\end{equation}

Here, \(Z_{i}\) shrinks slower than \(y_{i}\) as \(\lambda\) increases.

This implies
\(\frac{\partial}{\partial\lambda}\left( \frac{y_{i}}{Z_{i}} \right) < 0\)

\(\frac{\partial}{\partial\lambda}\left( - \frac{y_{i}}{Z_{i}} \right) > 0\).

Therefore, \(\frac{{\partial\tau}_{i}^{*}}{\partial\lambda} > 0\)

In the same way, the conditions for
\(\frac{{\partial\tau}_{i}^{*}}{\partial\gamma} > 0\) can be calculated
by writing out their expansions. First, we analyze
\(\frac{{\partial\tau}_{i}^{*}}{\partial\gamma}\).

\begin{equation}
    \begin{aligned}
        \frac{{\partial\tau}_{i}^{*}}{\partial\gamma} &= \frac{\partial}{\partial\gamma}\left( \frac{b + a\left( 1 - \frac{{2y}_{i}}{Z_{i}} \right)}{ab} \right)^{0.5} \\
        &= \left( \frac{a}{ab} \right)^{0.5}\frac{\partial}{\partial\gamma}{\left(\frac{b}{a} + 1 - \frac{{2y}_{i}}{Z_{i}}\right)}^{0.5}
    \end{aligned}
\end{equation}

We assume that \(\tau_{i}^{*}\) is defined, so we will assume
\(\frac{b}{a} + 1 - \frac{{2y}_{i}}{Z_{i}} > 0\)

Since \(\left( \frac{a}{ab} \right)^{0.5}\ \)is always positive, the
conditions for \(\frac{{\partial\tau}_{i}^{*}}{\partial\gamma}\) can be
simplified down to

\begin{equation}
    \frac{{\partial\tau}_{i}^{*}}{\partial\gamma} = \frac{\partial}{\partial\gamma}{\left(\frac{b}{a} + 1 - \frac{{2y}_{i}}{Z_{i}}\right)}^{0.5} = \frac{2*\frac{\partial}{\partial\gamma}\left( - \frac{y_{i}}{Z_{i}}\right)}{2{\left(\frac{b}{a} + 1 - \frac{{2y}_{i}}{Z_{i}}\right)}^{0.5}}
\end{equation}

We note that the denominator is positive, meaning the direction of
\(\frac{{\partial\tau}_{i}^{*}}{\partial\gamma}\) can be determined by
the direction and conditions of equation

\begin{equation}
    \frac{\partial}{\partial\gamma}\left( - \frac{y_{i}}{Z_{i}}\right)
\end{equation}

Since \(y_{i}\) has no \(\gamma\) term, we are left evaluating equation

\begin{equation}
    {- y}_{i}\frac{\partial}{\partial\gamma}\left( \frac{1}{Z_{i}} \right) > 0
\end{equation}

\begin{equation}
    {- y}_{i}\frac{Z_{i}\frac{\partial}{\partial\gamma}(1) - (1)\frac{\partial}{\partial\gamma}(Z_{i})}{{(Z_{i})}^{2}} > 0
\end{equation}

\begin{equation}
    \frac{\partial}{\partial\gamma}\left( Z_{i} \right) > 0
\end{equation}

We know that
\({(\gamma - \lambda)Z}_{i} = e^{(\gamma - \lambda)T_{i}} - e^{(\gamma - \lambda)T_{i - 1}}\),
so we calculate that

\begin{equation}
    \frac{\partial}{\partial\gamma}\left( (\gamma - \lambda)Z_{i} \right) = T_{i}e^{(\gamma - \lambda)T_{i}} - T_{i - 1}e^{(\gamma - \lambda)T_{i - 1}}
\end{equation}

\begin{equation}
    Z_{i}\frac{\partial}{\partial\gamma}(\gamma - \lambda) + (\gamma - \lambda)\frac{\partial}{\partial\gamma}Z_{i} = T_{i}e^{(\gamma - \lambda)T_{i}} - T_{i - 1}e^{(\gamma - \lambda)T_{i - 1}}
\end{equation}

\begin{equation}
    \frac{dZ_{i}}{d\gamma} = \frac{T_{i}e^{(\gamma - \lambda)T_{i}} - T_{i - 1}e^{(\gamma - \lambda)T_{i - 1}} - Z_{i}}{(\gamma - \lambda)}
\end{equation}

From this, we can see \(\frac{dZ_{i}}{d\gamma} > 0\) is true when
\(\gamma > \lambda\ and\ \gamma > 0\). Hence
\(\frac{{\partial\tau}_{i}^{*}}{\partial\gamma} > 0.\)

\end{appendixproof}

\begin{appendixproof}{COROLLARY~\ref{cor5}}

We begin with the expression for the optimal threshold signature timelock:

\begin{equation}
    T^{*} = \frac{1}{\gamma}\ln\left( \frac{p\left( \tau_{i - 1}^{*} \right)q\left( \tau_{i - 1}^{*} \right) - p\left( \tau_{i}^{*} \right)q\left( \tau_{i}^{*} \right)}{2\left( p\left( \tau_{i - 1}^{*} \right) - p\left( \tau_{i}^{*} \right) \right)} \right)
\end{equation}

Let us define:

\begin{equation}
    g(\lambda) = p\left( \tau_{i - 1}^{*} \right)q\left( \tau_{i - 1}^{*} \right) - p\left( \tau_{i}^{*} \right)q\left( \tau_{i}^{*} \right)
\end{equation}

\begin{equation}
    \begin{aligned}
        f(\lambda) &= 2\left( p\left( \tau_{i - 1}^{*} \right) - p\left( \tau_{i}^{*} \right) \right) \\
        &= 2a\left( {\tau_{i}^{*}}^{2} - {\tau_{i - 1}^{*}}^{2} \right)
    \end{aligned}
\end{equation}

Given that \(\tau_{i} < \tau_{i - 1}\), we have
\(\tau_{i}^{2} < \tau_{i - 1}^{2}\), so \(f(\lambda) < 0\). For
\(T^{*}\) to be valid, we also have \(g(\lambda) < 0\).

Substituting these definitions, we can rewrite:

\begin{equation}
    T^{*} = \frac{1}{\gamma}\ln\left( \frac{g(\lambda)}{f(\lambda)} \right)
\end{equation}

Taking the derivative with respect to \(\lambda\):

\begin{align}
    \frac{{\partial T}^{*}}{\partial\lambda} &= \frac{1}{\gamma} \cdot \frac{f(\lambda)g^{\prime}(\lambda) - g(\lambda)f^{\prime}(\lambda)}{g(\lambda)f(\lambda)} \\
    &= \frac{1}{\gamma}\left( \frac{g^{\prime}(\lambda)}{g(\lambda)} - \frac{f^{\prime}(\lambda)}{f(\lambda)} \right)
\end{align}

Since \(f\) satisfies the MLRP with respect to \(g\), the ratio
\(\frac{f(\lambda)}{g(\lambda)}\) is a monotonically increasing function
of \(\lambda\):

\begin{equation}
    \frac{d}{d\lambda}\left( \frac{f(\lambda)}{g(\lambda)} \right) > 0
\end{equation}

This derivative can be expanded as:

\begin{equation}
    \frac{d}{d\lambda}\left( \frac{f(\lambda)}{g(\lambda)} \right) = \frac{f^{\prime}(\lambda)g(\lambda) - f(\lambda)g^{\prime}(\lambda)}{\left\lbrack g(\lambda) \right\rbrack^{2}}
\end{equation}

Since \(g(\lambda) < 0\), we have
\(\left\lbrack g(\lambda) \right\rbrack^{2} > 0\), which means the sign
of \(\frac{d}{d\lambda}\left( \frac{f(\lambda)}{g(\lambda)} \right)\) is
determined by the sign of
\(\left( f^{\prime}(\lambda)g(\lambda) - f(\lambda)g^{\prime}(\lambda) \right)\).

Given that
\(\frac{d}{d\lambda}\left( \frac{f(\lambda)}{g(\lambda)} \right) > 0\)
due to MLRP, we must have:

\begin{equation}
    \left( f^{\prime}(\lambda)g(\lambda) - f(\lambda)g^{\prime}(\lambda) \right) > 0
\end{equation}

Dividing both sides by \(g(\lambda)f(\lambda)\) (which is positive since
both \(g(\lambda) < 0\) and \(f(\lambda) < 0\):

\begin{equation}
    \frac{f^{\prime}(\lambda)}{f(\lambda)} - \frac{g^{\prime}(\lambda)}{g(\lambda)} > 0
\end{equation}

Now returning to the expression for
\(\frac{{\partial T}^{*}}{\partial\lambda}\):

\begin{equation}
    \frac{{\partial T}^{*}}{\partial\lambda} = \frac{1}{\gamma}\left( \frac{g^{\prime}(\lambda)}{g(\lambda)} - \frac{f^{\prime}(\lambda)}{f(\lambda)} \right)
\end{equation}

Since
\(\frac{g^{\prime}(\lambda)}{g(\lambda)} - \frac{f^{\prime}(\lambda)}{f(\lambda)} < 0\)
and \(\gamma\) is positive, we have:

\begin{equation}
    \frac{{\partial T}^{*}}{\partial\lambda} < 0
\end{equation}

Therefore, the optimal threshold signature timelock \(T^{*}\) decreases as
\(\lambda\) increases.

We begin with the expression for the optimal threshold signature timelock:

\begin{equation}
    T^{*} = \frac{1}{\gamma}\ln\left( \frac{p\left( \tau_{i - 1}^{*} \right)q\left( \tau_{i - 1}^{*} \right) - p\left( \tau_{i}^{*} \right)q\left( \tau_{i}^{*} \right)}{2\left( p\left( \tau_{i - 1}^{*} \right) - p\left( \tau_{i}^{*} \right) \right)} \right)
\end{equation}

Let us define:

\begin{equation}
    g(\gamma) = p\left( \tau_{i - 1}^{*} \right)q\left( \tau_{i - 1}^{*} \right) - p\left( \tau_{i}^{*} \right)q\left( \tau_{i}^{*} \right)
\end{equation}

\begin{equation}
    \begin{aligned}
        f(\gamma) &= 2\left( p\left( \tau_{i - 1}^{*} \right) - p\left( \tau_{i}^{*} \right) \right) \\
        &= 2a\left( {\tau_{i}^{*}}^{2} - {\tau_{i - 1}^{*}}^{2} \right)
    \end{aligned}
\end{equation}

Given that \(\tau_{i} < \tau_{i - 1}\), we have
\(\tau_{i}^{2} < \tau_{i - 1}^{2}\), so \(f(\gamma) < 0\). For \(T^{*}\)
to be valid, we also have \(g(\gamma) < 0\).

Substituting these definitions, we can rewrite:

\begin{equation}
    T^{*} = \frac{1}{\gamma}\ln\left( \frac{g(\gamma)}{f(\gamma)} \right)
\end{equation}

Taking the derivative with respect to \(\gamma\):

\begin{align}
    \frac{{\partial T}^{*}}{\partial\gamma} &= - \frac{1}{\gamma^{2}}\ln\left( \frac{g(\gamma)}{f(\gamma)} \right) + \frac{1}{\gamma} \nonumber \\
    &\hspace{10pt} \cdot \frac{f(\gamma)g^{\prime}(\gamma) - g(\gamma)f^{\prime}(\gamma)}{g(\gamma)f(\gamma)} \\
    &= - \frac{T^{*}}{\gamma} + \frac{1}{\gamma}\left( \frac{g^{\prime}(\gamma)}{g(\gamma)} - \frac{f^{\prime}(\gamma)}{f(\gamma)} \right)
\end{align}

Since \(f\) satisfies the MLRP with respect to \(g\), the ratio
\(\frac{f(\gamma)}{g(\gamma)}\) is a monotonically increasing function
of \(\gamma\):

\begin{equation}
    \frac{d}{d\gamma}\left( \frac{f(\gamma)}{g(\gamma)} \right) > 0
\end{equation}

This derivative can be expanded as:

\begin{equation}
    \frac{d}{d\gamma}\left( \frac{f(\gamma)}{g(\gamma)} \right) = \frac{f^{\prime}(\gamma)g(\gamma) - f(\gamma)g^{\prime}(\gamma)}{\left\lbrack g(\gamma) \right\rbrack^{2}}
\end{equation}

Since \(g(\gamma) < 0\), we have
\(\left\lbrack g(\gamma) \right\rbrack^{2} > 0\), which means the sign
of \(\frac{d}{d\gamma}\left( \frac{f(\gamma)}{g(\gamma)} \right)\) is
determined by the sign of
\(\left( f^{\prime}(\gamma)g(\gamma) - f(\gamma)g^{\prime}(\gamma) \right)\).

Given that
\(\frac{d}{d\gamma}\left( \frac{f(\gamma)}{g(\gamma)} \right) > 0\) due
to MLRP, we must have:

\begin{equation}
    \left( f^{\prime}(\gamma)g(\gamma) - f(\gamma)g^{\prime}(\gamma) \right) > 0
\end{equation}

Dividing both sides by \(g(\gamma)f(\gamma)\) (which is positive since
both \(g(\gamma) < 0\) and \(f(\gamma) < 0\):

\begin{align}
    \frac{f^{\prime}(\gamma)}{f(\gamma)} - \frac{g^{\prime}(\gamma)}{g(\gamma)} > 0 \\
    \frac{g^{\prime}(\gamma)}{g(\gamma)} - \frac{f^{\prime}(\gamma)}{f(\gamma)} < 0
\end{align}

Now returning to the expression for
\(\frac{{\partial T}^{*}}{\partial\gamma}\):

\begin{equation}
    \frac{{\partial T}^{*}}{\partial\gamma} = - \frac{T^{*}}{\gamma} + \frac{1}{\gamma}\left( \frac{g^{\prime}(\gamma)}{g(\gamma)} - \frac{f^{\prime}(\gamma)}{f(\gamma)} \right)
\end{equation}

Since
\(\frac{g^{\prime}(\gamma)}{g(\gamma)} - \frac{f^{\prime}(\gamma)}{f(\gamma)} < 0\)
and \(- \frac{T^{*}}{\gamma} < 0\) (as both \(T^{*}\) and \(\gamma\) are
positive), we have:

\begin{equation}
    \frac{{\partial T}^{*}}{\partial\gamma} < 0
\end{equation}

Therefore, the optimal threshold signature timelock \(T^{*}\) decreases as
\(\gamma\) increases.

\end{appendixproof}

\begin{appendixproof}{COROLLARY~\ref{cor9}}

We begin with the expression for the optimal threshold signature timelock:

\begin{equation}
    T^{*} = \frac{1}{\gamma}\ln\left( \frac{2\left( p\left( \tau_{i - 1}^{*} \right) - p\left( \tau_{i}^{*} \right) \right)}{p\left( \tau_{i - 1}^{*} \right)q\left( \tau_{i - 1}^{*} \right) - p\left( \tau_{i}^{*} \right)q\left( \tau_{i}^{*} \right)} \right)
\end{equation}

Let us define:

\begin{align}
    g(\lambda) &= p\left( \tau_{i - 1}^{*} \right)q\left( \tau_{i - 1}^{*} \right) - p\left( \tau_{i}^{*} \right)q\left( \tau_{i}^{*} \right) \\
    f(\lambda) &= 2\left( p\left( \tau_{i - 1}^{*} \right) - p\left( \tau_{i}^{*} \right) \right) = 2a\left( {\tau_{i}^{*}}^{2} - {\tau_{i - 1}^{*}}^{2} \right)
\end{align}

Given that \(\tau_{i} < \tau_{i - 1}\), we have
\(\tau_{i}^{2} < \tau_{i - 1}^{2}\), so \(f(\lambda) < 0\). For
\(T^{*}\) to be valid, we also have \(g(\lambda) < 0\).

Substituting these definitions, we can rewrite:

\begin{equation}
    T^{*} = \frac{1}{\gamma}\ln\left( \frac{f(\lambda)}{g(\lambda)} \right)
\end{equation}

Taking the derivative with respect to \(\lambda\):

\begin{align}
    \frac{{\partial T}^{*}}{\partial\lambda} &= \frac{1}{\gamma} \cdot \frac{g(\lambda)f^{\prime}(\lambda) - f(\lambda)g^{\prime}(\lambda)}{g(\lambda)f(\lambda)} \\
    &= \frac{1}{\gamma}\left( \frac{f^{\prime}(\lambda)}{f(\lambda)} - \frac{g^{\prime}(\lambda)}{g(\lambda)} \right)
\end{align}

Since \(f\) satisfies the MLRP with respect to \(g\), the ratio
\(\frac{g(\lambda)}{f(\lambda)}\) is a monotonically decreasing function
of \(\lambda\):

\begin{equation}
    \frac{d}{d\lambda}\left( \frac{g(\lambda)}{f(\lambda)} \right) < 0
\end{equation}

This derivative can be expanded as:

\begin{equation}
    \frac{d}{d\lambda}\left( \frac{g(\lambda)}{f(\lambda)} \right) = \frac{g^{\prime}(\lambda)f(\lambda) - g(\lambda)f^{\prime}(\lambda)}{\left\lbrack f(\lambda) \right\rbrack^{2}}
\end{equation}

Since $f(\lambda) < 0$, we have $\left\lbrack f(\lambda) \right\rbrack^{2} > 0$, which means the sign of $\frac{d}{d\lambda}\left( \frac{g(\lambda)}{f(\lambda)} \right)$ is determined by the sign of $\left( g^{\prime}(\lambda)f(\lambda) - g(\lambda)f^{\prime}(\lambda) \right)$.

Given that
\(\frac{d}{d\lambda}\left( \frac{g(\lambda)}{f(\lambda)} \right) < 0\)
due to MLRP, we must have:

\begin{equation}
    \left( f(\lambda)g^{\prime}(\lambda) - f^{\prime}(\lambda)g(\lambda) \right) < 0
\end{equation}

Dividing both sides by \(g(\lambda)f(\lambda)\) (which is positive since
both \(g(\lambda) < 0\) and \(f(\lambda) < 0\):

\begin{align}
    \frac{g^{\prime}(\lambda)}{g(\lambda)} - \frac{f^{\prime}(\lambda)}{f(\lambda)} < 0 \\
    \frac{f^{\prime}(\lambda)}{f(\lambda)} - \frac{g^{\prime}(\lambda)}{g(\lambda)} > 0
\end{align}

Now returning to the expression for
\(\frac{{\partial T}^{*}}{\partial\lambda}\):

\begin{equation}
    \frac{{\partial T}^{*}}{\partial\lambda} = \frac{1}{\gamma}\left( \frac{f^{\prime}(\lambda)}{f(\lambda)} - \frac{g^{\prime}(\lambda)}{g(\lambda)} \right)
\end{equation}

Since
\(\frac{f^{\prime}(\lambda)}{f(\lambda)} - \frac{g^{\prime}(\lambda)}{g(\lambda)} > 0\)
and \(\gamma\) is positive, we have:

\begin{equation}
    \frac{{\partial T}^{*}}{\partial\lambda} > 0
\end{equation}

Therefore, the optimal threshold signature timelock \(T^{*}\) increases as
\(\lambda\) increases.

We begin with the expression for the optimal threshold signature timelock:

\begin{equation}
    T^{*} = \frac{1}{\gamma}\ln\left( \frac{2\left( p\left( \tau_{i - 1}^{*} \right) - p\left( \tau_{i}^{*} \right) \right)}{p\left( \tau_{i - 1}^{*} \right)q\left( \tau_{i - 1}^{*} \right) - p\left( \tau_{i}^{*} \right)q\left( \tau_{i}^{*} \right)} \right)
\end{equation}

Let us define:

\begin{align}
    g(\gamma) &= p\left( \tau_{i - 1}^{*} \right)q\left( \tau_{i - 1}^{*} \right) - p\left( \tau_{i}^{*} \right)q\left( \tau_{i}^{*} \right) \\
    f(\lambda) &= 2\left( p\left( \tau_{i - 1}^{*} \right) - p\left( \tau_{i}^{*} \right) \right) = 2a\left( {\tau_{i}^{*}}^{2} - {\tau_{i - 1}^{*}}^{2} \right)
\end{align}

Given that \(\tau_{i} < \tau_{i - 1}\), we have
\(\tau_{i}^{2} < \tau_{i - 1}^{2}\), so \(f(\gamma) < 0\). For \(T^{*}\)
to be valid, we also have \(g(\gamma) < 0\).

Substituting these definitions, we can rewrite:

\begin{equation}
    T^{*} = \frac{1}{\gamma}\ln\left( \frac{f(\gamma)}{g(\gamma)} \right)
\end{equation}

Taking the derivative with respect to $\gamma$:

\begin{align}
    \frac{{\partial T}^{*}}{\partial\gamma} &= - \frac{1}{\gamma^{2}}\ln\left( \frac{f(\gamma)}{g(\gamma)} \right) + \frac{1}{\gamma} \nonumber \\
    &\hspace{10pt} \cdot \frac{g(\gamma)f^{\prime}(\gamma) - f(\gamma)g^{\prime}(\gamma)}{g(\gamma)f(\gamma)} \\
    &= - \frac{T^{*}}{\gamma} + \frac{1}{\gamma}\left( \frac{f^{\prime}(\gamma)}{f(\gamma)} - \frac{g^{\prime}(\gamma)}{g(\gamma)} \right)
\end{align}

Since \(f\) satisfies the MLRP with respect to \(g\), the ratio
\(\frac{g(\gamma)}{f(\gamma)}\) is a monotonically decreasing function
of \(\gamma\):

\begin{equation}
    \frac{d}{d\gamma}\left( \frac{g(\gamma)}{f(\gamma)} \right) < 0
\end{equation}

This derivative can be expanded as:

\begin{equation}
    \frac{d}{d\lambda}\left( \frac{g(\gamma)}{f(\gamma)} \right) = \frac{g^{\prime}(\gamma)f(\gamma) - g(\gamma)f^{\prime}(\gamma)}{\left\lbrack f(\gamma) \right\rbrack^{2}}
\end{equation}

Since $f(\lambda) < 0$, we have $\left\lbrack f(\lambda) \right\rbrack^{2} > 0$, which means the sign of $\frac{d}{d\lambda}\left( \frac{g(\lambda)}{f(\lambda)} \right)$ is determined by the sign of $\left( g^{\prime}(\lambda)f(\lambda) - g(\lambda)f^{\prime}(\lambda) \right)$.

Given that
\(\frac{d}{d\lambda}\left( \frac{g(\lambda)}{f(\lambda)} \right) < 0\)
due to MLRP, we must have:

\begin{equation}
    \left( f(\lambda)g^{\prime}(\lambda) - f^{\prime}(\lambda)g(\lambda) \right) < 0
\end{equation}

Dividing both sides by \(g(\lambda)f(\lambda)\) (which is positive since
both \(g(\lambda) < 0\) and \(f(\lambda) < 0\):

\begin{align}
    \frac{g^{\prime}(\lambda)}{g(\lambda)} - \frac{f^{\prime}(\lambda)}{f(\lambda)} < 0 \\
    \frac{f^{\prime}(\lambda)}{f(\lambda)} - \frac{g^{\prime}(\lambda)}{g(\lambda)} > 0
\end{align}

Now returning to the expression for
\(\frac{{\partial T}^{*}}{\partial\lambda}\):

\begin{equation}
    \frac{{\partial T}^{*}}{\partial\lambda} = \frac{1}{\gamma}\left( \frac{f^{\prime}(\lambda)}{f(\lambda)} - \frac{g^{\prime}(\lambda)}{g(\lambda)} \right)
\end{equation}

Since
\(\frac{f^{\prime}(\lambda)}{f(\lambda)} - \frac{g^{\prime}(\lambda)}{g(\lambda)} > 0\)
and \(\gamma\) is positive, we have:

\begin{equation}
    \frac{{\partial T}^{*}}{\partial\lambda} > 0
\end{equation}

Therefore, the optimal threshold signature timelock \(T^{*}\) increases as
\(\lambda\) increases.

We begin with the expression for the optimal threshold signature timelock:

\begin{equation}
    T^{*} = \frac{1}{\gamma}\ln\left( \frac{2\left( p\left( \tau_{i - 1}^{*} \right) - p\left( \tau_{i}^{*} \right) \right)}{p\left( \tau_{i - 1}^{*} \right)q\left( \tau_{i - 1}^{*} \right) - p\left( \tau_{i}^{*} \right)q\left( \tau_{i}^{*} \right)} \right)
\end{equation}

Let us define:

\begin{align}
    g(\gamma) &= p\left( \tau_{i - 1}^{*} \right)q\left( \tau_{i - 1}^{*} \right) - p\left( \tau_{i}^{*} \right)q\left( \tau_{i}^{*} \right) \\
    f(\lambda) &= 2\left( p\left( \tau_{i - 1}^{*} \right) - p\left( \tau_{i}^{*} \right) \right) = 2a\left( {\tau_{i}^{*}}^{2} - {\tau_{i - 1}^{*}}^{2} \right)
\end{align}

Given that \(\tau_{i} < \tau_{i - 1}\), we have
\(\tau_{i}^{2} < \tau_{i - 1}^{2}\), so \(f(\gamma) < 0\). For \(T^{*}\)
to be valid, we also have \(g(\gamma) < 0\).

Substituting these definitions, we can rewrite:

\begin{equation}
    T^{*} = \frac{1}{\gamma}\ln\left( \frac{f(\gamma)}{g(\gamma)} \right)
\end{equation}

Taking the derivative with respect to $\gamma$:

\begin{align}
    \frac{{\partial T}^{*}}{\partial\gamma} &= - \frac{1}{\gamma^{2}}\ln\left( \frac{f(\gamma)}{g(\gamma)} \right) + \frac{1}{\gamma} \nonumber \\
    &\hspace{10pt} \cdot \frac{g(\gamma)f^{\prime}(\gamma) - f(\gamma)g^{\prime}(\gamma)}{g(\gamma)f(\gamma)} \\
    &= - \frac{T^{*}}{\gamma} + \frac{1}{\gamma}\left( \frac{f^{\prime}(\gamma)}{f(\gamma)} - \frac{g^{\prime}(\gamma)}{g(\gamma)} \right)
\end{align}

Since \(f\) satisfies the MLRP with respect to \(g\), the ratio
\(\frac{g(\gamma)}{f(\gamma)}\) is a monotonically decreasing function
of \(\gamma\):

\begin{equation}
    \frac{d}{d\gamma}\left( \frac{g(\gamma)}{f(\gamma)} \right) < 0
\end{equation}

This derivative can be expanded as:

\begin{equation}
    \frac{d}{d\lambda}\left( \frac{g(\gamma)}{f(\gamma)} \right) = \frac{g^{\prime}(\gamma)f(\gamma) - g(\gamma)f^{\prime}(\gamma)}{\left\lbrack f(\gamma) \right\rbrack^{2}}
\end{equation}

Since \(f(\gamma) < 0\), we have
\(\left\lbrack f(\gamma) \right\rbrack^{2} > 0\), which means the sign
of \(\frac{d}{d\gamma}\left( \frac{g(\gamma)}{f(\gamma)} \right)\) is
determined by the sign of
\(g^{\prime}(\gamma)f(\gamma) - g(\gamma)f^{\prime}(\gamma)\).

Given that
\(\frac{d}{d\gamma}\left( \frac{g(\gamma)}{f(\gamma)} \right) < 0\) due
to MLRP, we must have:

\begin{equation}
    \left( g^{\prime}(\gamma)f(\gamma) - g(\gamma)f^{\prime}(\gamma) \right) < 0
\end{equation}

Dividing both sides by \(g(\gamma)f(\gamma)\) (which is positive since
both \(g(\gamma) < 0\) and \(f(\gamma) < 0\):

\begin{align}
    \frac{g^{\prime}(\gamma)}{g(\gamma)} - \frac{f^{\prime}(\gamma)}{f(\gamma)} < 0 \\
    \frac{f^{\prime}(\gamma)}{f(\gamma)} - \frac{g^{\prime}(\gamma)}{g(\gamma)} > 0
\end{align}

Now returning to the expression for
\(\frac{{\partial T}^{*}}{\partial\gamma}\):

\begin{equation}
    \frac{{\partial T}^{*}}{\partial\gamma} = - \frac{T^{*}}{\gamma} + \frac{1}{\gamma}\left( \frac{f^{\prime}(\gamma)}{f(\gamma)} - \frac{g^{\prime}(\gamma)}{g(\gamma)} \right)
\end{equation}

Since
\(\left( \frac{f^{\prime}(\gamma)}{f(\gamma)} - \frac{g^{\prime}(\gamma)}{g(\gamma)} \right) > 0\)
and \(- \frac{T^{*}}{\gamma} < 0\) (as both \(T^{*}\) and \(\gamma\) are
positive), we have:

\begin{equation}
    \frac{{\partial T}^{*}}{\partial\gamma} > 0 \ \text{ if } \ \frac{f^{\prime}(\gamma)}{f(\gamma)} - \frac{g^{\prime}(\gamma)}{g(\gamma)}
\end{equation}

Therefore, the optimal threshold signature timelock \(T^{*}\) decreases as
\(\gamma\) increases.
\end{appendixproof}

\end{document}